\documentclass[sn-mathphys, Numbered]{sn-jnl}% Math and Physical Sciences Reference Style
\usepackage{graphicx}%
\usepackage{multirow}%
\usepackage{amsmath,amssymb,amsfonts}%
\usepackage{amsthm}%
\usepackage{mathrsfs}%
\usepackage[title]{appendix}%
\usepackage{xcolor}%
\usepackage{textcomp}%
\usepackage{manyfoot}%
\usepackage{booktabs}%
\usepackage{algorithm}%
\usepackage{algorithmicx}%
\usepackage{algpseudocode}%
\usepackage{listings}%
\usepackage{colortbl}
\usepackage{booktabs}
\usepackage[normalem]{ulem}
\usepackage{moresize}
\usepackage{amsmath,amssymb,amsfonts}
\usepackage{times}  
\usepackage{titlesec}
\usepackage{array, multirow,graphicx}
\usepackage{subfigure}
\usepackage[export]{adjustbox}
\begin{document}

\title[Article Title]{iRED: A disaggregated P4-AQM fully implemented in programmable data plane hardware}

%%=============================================================%%
%% Prefix	-> \pfx{Dr}
%% GivenName	-> \fnm{Joergen W.}
%% Particle	-> \spfx{van der} -> surname prefix
%% FamilyName	-> \sur{Ploeg}
%% Suffix	-> \sfx{IV}
%% NatureName	-> \tanm{Poet Laureate} -> Title after name
%% Degrees	-> \dgr{MSc, PhD}
%% \author*[1,2]{\pfx{Dr} \fnm{Joergen W.} \spfx{van der} \sur{Ploeg} \sfx{IV} \tanm{Poet Laureate} 
%%                 \dgr{MSc, PhD}}\email{iauthor@gmail.com}
%%=============================================================%%

\author*[1,4]{\fnm{Leandro} \sur{C. de Almeida}}\email{leandro.almeida@ifpb.edu.br}

\author[2]{\fnm{Rafael} \sur{Pasquini}}\email{rafael.pasquini@ufu.br}
\equalcont{These authors contributed equally to this work.}

\author[3]{\fnm{Chrysa} \sur{Papagianni}}\email{c.papagianni@uva.nl}
\equalcont{These authors contributed equally to this work.}

\author[4]{\fnm{Fábio} \sur{L. Verdi}}\email{verdi@ufscar.br}
\equalcont{These authors contributed equally to this work.}

\affil*[1]{\orgdiv{Academic Unit of Informatics}, \orgname{Federal Institute of Paraíba}, \orgaddress{\street{Av. Primeiro de Maio}, \city{João Pessoa}, \postcode{58015-435}, \state{PB}, \country{Brazil}}}

\affil[2]{\orgdiv{Faculty of Computing}, \orgname{Federal University of Uberlândia}, \orgaddress{\street{Av. João Naves de Ávila}, \city{Uberlândia}, \postcode{38400-902}, \state{MG}, \country{Brazil}}}

\affil[3]{\orgdiv{Faculty of Science}, \orgname{University of Amsterdam}, \orgaddress{\street{1098 XH Amsterdam}, \country{Netherlands}}}

\affil[4]{\orgdiv{Department of Computer Science}, \orgname{Federal University of São Carlos - Sorocaba Campus}, \orgaddress{\street{Rodovia João Leme dos Santos}, \city{Sorocaba}, \postcode{18052-780}, \state{SP}, \country{Brazil}}}

\abstract{Routers employ queues to temporarily hold packets when the scheduler cannot immediately process them. Congestion occurs when the arrival rate of packets exceeds the processing capacity, leading to increased queueing delay. Over time, Active Queue Management (AQM) strategies have focused on directly draining packets from queues to alleviate congestion and reduce queuing delay. On Programmable Data Plane (PDP) hardware, AQMs traditionally reside in the Egress pipeline due to the availability of queue delay information there. We argue that this approach wastes the router's resources because the dropped packet has already consumed the entire pipeline of the device. In this work, we propose ingress Random Early Detection (iRED), a more efficient approach that addresses the Egress drop problem. iRED is a disaggregated P4-AQM fully implemented in programmable data plane hardware and also supports Low Latency, Low Loss, and Scalable Throughput (L4S) framework, saving device pipeline resources by dropping packets in the Ingress block. To evaluate iRED, we conducted three experiments using a Tofino2 programmable switch: $i)$ An in-depth analysis of state-of-the-art AQMs on PDP hardware, using 12 different network configurations varying in bandwidth, Round-Trip Time (RTT), and Maximum Transmission Unit (MTU). The results demonstrate that iRED can significantly reduce router resource consumption, with up to a 10x reduction in memory usage, 12x fewer processing cycles, and 8x less power consumption for the same traffic load; $ii)$ A performance evaluation regarding the L4S framework. The results prove that iRED achieves fairness in bandwidth usage for different types of traffic (classic and scalable); $iii)$ A comprehensive analysis of the Quality of Service (QoS) in a real setup of a Dynamic Adaptive Streaming over HTTP (DASH) technology. iRED demonstrated up to a 2.34x improvement in Frames Per Second (FPS) and a 4.77x increase in the video player buffer fill.}


%%\pacs[JEL Classification]{D8, H51}

%%\pacs[MSC Classification]{35A01, 65L10, 65L12, 65L20, 65L70}

\maketitle

\section{Introduction}\label{sec:intro}

High bandwidth and low latency are key features of current applications (e.g., video streaming, cloud games, telesurgery, and others) that run on modern communication networks.  To meet the strict requirements of each application, routers need to accommodate the large volume of traffic generated by users. When the packet arrival rate exceeds the processing capacity, they are accommodated temporarily in the appropriate output queue, likely causing packet delay. In this case, users of applications sensitive to the delay tend to suffer, as this delay reduces the Quality of Service (QoS) delivered.

There have been strategies to deal with this problem since the beginning of the Internet, such as congestion control (CC) mechanisms \cite{cc:88}. CC continuously monitors current connections, allowing dynamic adjustment of network segment sending rates. In other words, it manages when a host should increase or decrease its transmission rate, trying to make better use of network resources. The Transmission Control Protocol (TCP) \cite{rfc793} is, without a doubt, the primary protocol that implemented the CC mechanism by end hosts in recent decades.

However, the TCP algorithm needs to receive feedback on the network state, which can come in the form of a congestion signal. The primary methods for conveying congestion conditions to senders include packet \textbf{marking} using Explicit Congestion Notification (ECN) bits and selective packet \textbf{dropping}. Active Queue Management (AQM) is a traditional mechanism employed in network device queues, such as routers and switches, to assist the CC capable of implementing these two functions (mark and drop).

In this context, AQMs such as RED \cite{RED:93}, BLUE \cite{Blue:2002}, CoDel \cite{CODEL:2012}, CAKE \cite{CAKE:2018} and PIE \cite{ietf-aqm-pie-03} has been used to drop packets when the queue builds-up, alleviating the congestion and reducing the queueing delay. In a classical router, proposing and evaluating new strategies can be costly, since fundamental changes to the ASIC (such as a new AQM implementation) have traditionally required building a new ASIC involving costly hardware updates \cite{Tofino+P4}. Thanks to recent advances in data plane programmability and languages as Programming Protocol-independent Packet Processors (P4) \cite{P4-paper} and Network Programming Language (NPL) \cite{NPL}, it is possible to implement new AQMs functionality without having to redesign the ASIC.

In this direction, to the best of our knowledge, the prominent state-of-the-art AQMs implemented for running in PDP hardware and publicly available are P4-CoDel \cite{kundel2021p4codel} and (dual) PI2 \cite{DualPI2}. Although the approaches have different logic, both use the queueing delay information as input for the respective algorithm to decide whether or not to drop packets. After making the drop decision, the algorithm must set the current packet to be discarded for an effective drop action at the end of the pipeline. \textit{We argue that this approach wastes the router's resources, as the marked packet has already consumed the entire pipeline of the device}. In this work, we discuss this topic as \textbf{the Egress drop problem} in more depth in Section \ref{sec:keyconcepts}. 

\begin{figure}[ht]
    \centering
    \includegraphics[width=.7\columnwidth]{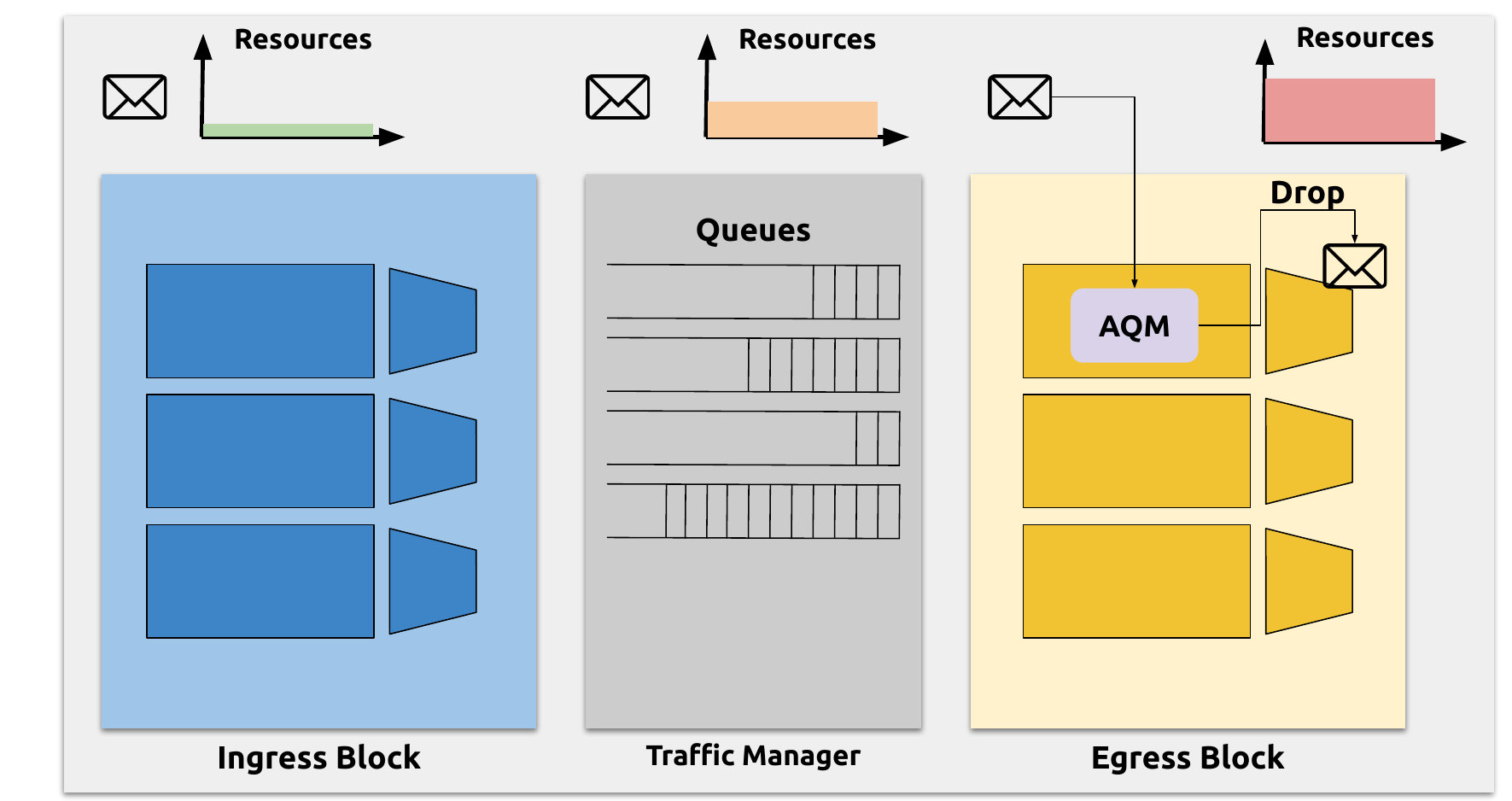}
    \caption{\textbf{The Egress drop problem}: When the AQM decision is implemented at the Egress, the packets dropped waste device resources.}
    \label{fig:problem}
\end{figure}

Fig. 1 illustrates the Egress drop problem in a generic PDP architecture, in which dropped packets consume device pipeline resources. Such architecture represents the current programmable switches architecture available in the market, such as the Tofino Native Architecture (TNA) \cite{TNA} and the Broadcom Trident4 / BCM56880 Series \cite{Broadcom}. 

Initially, incoming packets are received at the Ingress Block and the match-action logic (e.g., IPv4 forwarding) is executed. In the Traffic Manager, which is non-programmable, packets can be accommodated temporarily in the appropriate output queue. After this, packets are sent to the Egress Block, where AQMs - such as P4-CoDel and (dual) PI2 - algorithms are traditionally implemented as a match-action logic to make drop decisions (set a packet to drop). Finally, the packets marked to be dropped will be effectively discarded in Egress Deparser, which is the last phase in the pipeline. 

One may ask why the AQMs are deployed in the Egress block causing a waste of resources since the packet traversed all the pipelines in the switch and then was discarded. It makes more sense to deploy the AQMs in the Ingress block. However, queuing delay metadata (or queue depth) which is the main information used as input to the AQM algorithm to decide whether the packet should be dropped or not, is captured by the Traffic Manager and made available only in the Egress block. So, the challenge here is to design a solution in which the packets are dropped in the Ingress block saving resources of the network device. 

In addition to this context, recent efforts \cite{L4S:2023} by the Internet Engineering Task Force (IETF) have led to an architecture enabling Internet applications to achieve low queuing latency, low congestion loss, and scalable throughput control (L4S). The L4S architecture introduces incremental changes to both hosts and network nodes. On the host side, L4S incorporates a novel variant of a ``Scalable" CC algorithm known as TCP Prague \cite{prague:2018}. TCP Prague adjusts its window reduction in proportion to the extent of recently observed congestion. This stands in contrast to ``Classic" CC algorithms, which typically implement a worst-case reduction, typically by half, upon detecting any sign of congestion. At network nodes, L4S brings a dual queue coupled mechanism \cite{rfc9332}, in which one queue is for Classic traffic and another queue is for Scalable traffic. This coupled mechanism allows fair use of bandwidth, ensuring harmonious coexistence between CC flavors.

%The L4S architecture is based on the insight that the primary source of queuing delay resides within the behavior of the senders, as opposed to being an inherent property of the queuing system. One of the challenges inherent in this emerging architecture lies in the transition from classic CC algorithms to the assimilation of a novel category of CC algorithms adept at seeking network capacity while minimizing queuing latency. In this context, documents disseminated by the IETF \cite{L4S:2023} propose an incremental deployment strategy, facilitating the coexistence and equitable utilization of network bandwidth resources.}

In light of the Egress drop problem and a comprehensive understanding of the L4S architecture, in this work we propose a new approach denominated ingress Random Early Detection (iRED), to the best of our efforts, the only currently deployable P4-based AQM in the PDP that supports L4S. iRED splits the AQM logic into two parts: the decision and the action. The decision part, which depends on the queuing delay metadata, is deployed in the Egress block. The action part, which is responsible for dropping the packet, is deployed in the Ingress block. Additionally, it accomplishes this by categorizing traffic as either Classic (subject to dropping) or Scalable (marked with the ECN bit), thus ensuring fairness among various flows through a combined packet dropping and marking mechanism.

%Understanding that the \textbf{decision} to drop can be separated from the actual dropping \textbf{action}, we fully implemented a new disaggregated\footnote{Disaggregated in this work means performing the AQM operations in different blocks of the PDP architecture.} AQM mechanism named iRED (ingress Random Early Detection) in which packets are dropped at the Ingress block. iRED was implemented in P4 and runs in a tofino programmable hardware.

We conducted three experiments. First, an in-depth evaluation of the resource consumption of state-of-the-art AQMs available on PDP hardware was performed with 12 different network configurations, varying bandwidth, RTT, and MTU. Experiments show that our solution can offer a significant reduction in router resources, up to 10x less memory, 12x fewer cycles, and 8x less power for the same traffic load. Secondly, we conducted a comparative study between Classic and Scalable flows under non-stationary traffic conditions in an L4S environment. The results substantiate that iRED effectively ensures equitable utilization of bandwidth across various traffic types, including both classic and scalable. Finally, we assess the QoS of DASH in a real-world scenario. In this experiment, iRED exhibited a 2.34x improvement in FPS and a 4.77x increase in video player buffer fill.

In this paper, we present the following contributions:

\begin{enumerate}
    \item We investigate up-to-date research on AQM strategies implemented on the programmable data planes, identifying, characterizing, and clarifying the Egress drop problem.
    \item We design and implement iRED, a P4-AQM fully implemented in programmable data plane hardware also being L4S-capable, that introduces the concept of disaggregated AQM, effectively resolving the Egress drop problem.
    \item We conduct a comprehensive assessment of resource consumption by AQMs on a Tofino2 programmable switch. Our findings substantiate the premise that Ingress drop highly conserves switch resources.
\end{enumerate}

The remainder of the paper is structured as follows. In Section \ref{sec:keyconcepts}, we clarify the key concepts for a better understanding of the iRED. iRED is detailed in Section \ref{sec:iRED}. Additionally, we give a brief overview of state-of-the-art AQMs in Section \ref{sec:AQMOperation}. Evaluation and results are detailed, including a brief view of the testbed and the workloads used in Section \ref{sec:evaluation}. Finally, conclusions are drawn in Section \ref{sec:conclusions}.

\section{Key concepts} \label{sec:keyconcepts}

This section covers essential topics for a clear understanding of iRED. Initially, we will address the Egress drop problem in detail. Later, we will present a brief summary of the L4S framework.

\subsection{The Egress drop problem - A brief overview} \label{sec:problem}

In this subsection, we elucidate the operation of a typical programmable data plane switch, providing insights into the precise conditions that give rise to the Egress drop problem – where, how, and why it occurs. Additionally, we expound upon the advantages of separating the decision-making process from the packet-discarding action within an AQM logic. To comprehensively grasp the origins of this issue, we delve deeper into the architecture of a standard programmable switch.

\begin{figure}[ht]
    \centering
    \includegraphics[width=.8\columnwidth]{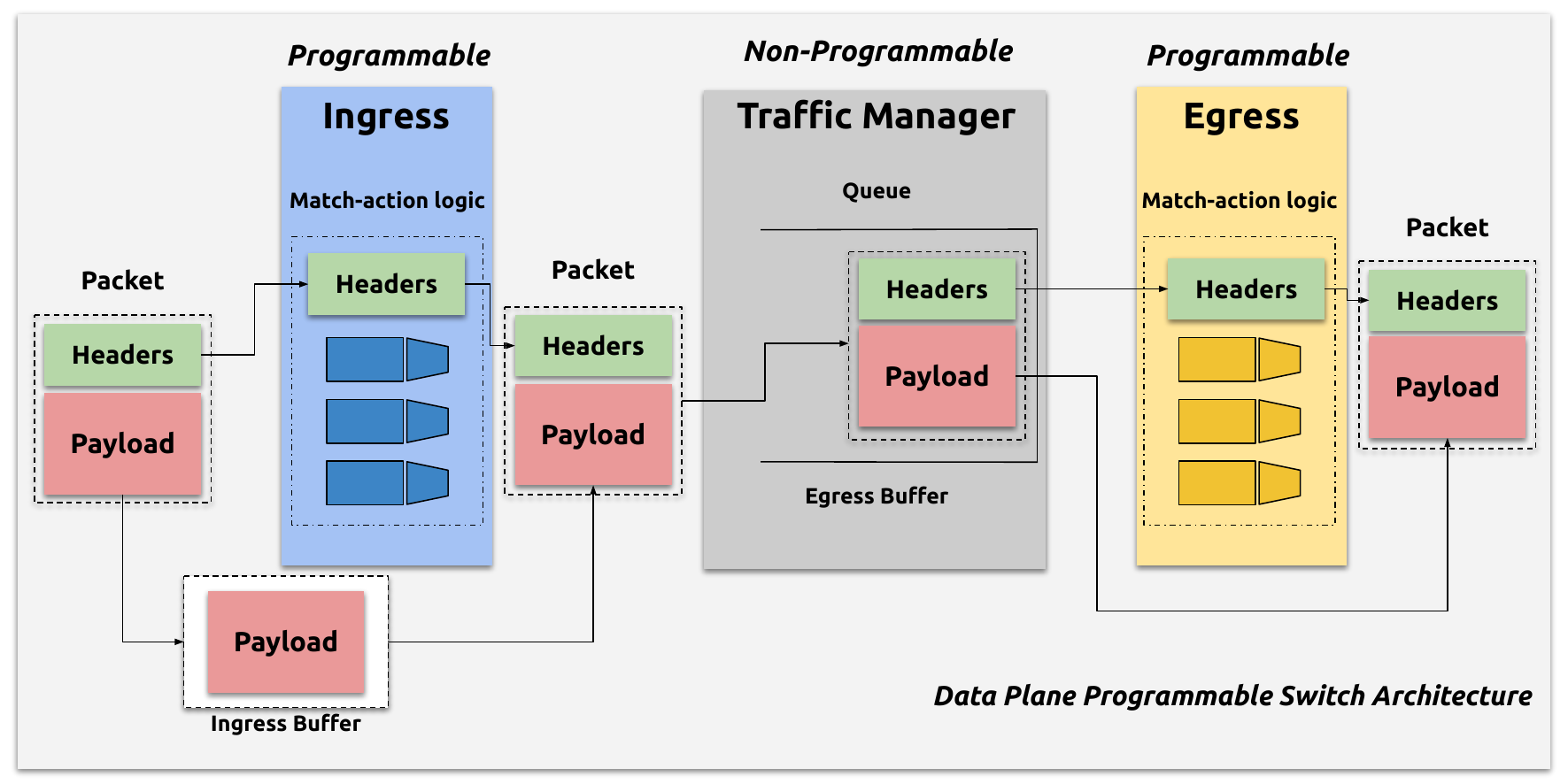}
    \caption{The generic architecture of the data plane programmable switch. Headers and Payload follow different paths on the device. }
    \label{fig:switcharch}
\end{figure}

As detailed in Fig. \ref{fig:switcharch}, a generic switch architecture with a programmable data plane is composed of some programmable blocks (Ingress and Egress) and non-programmable components (Traffic Manager). After the packet is received by a given ingress port, it is separated into Headers and Payload. Headers are the structures that are actually processed by programmable blocks. It is from the data contained in the header fields and other metadata that the programmer can define logic based on match-action to accomplish what is desired with the network packet. On the other hand, the Payload remains unchanged, usually stored in buffers, throughout the packet processing. After processing the Ingress block, the packet is reconstructed, generally by a Deparser (omitted in the Figure), which unifies the header with the payload that was stored in the Ingress buffer. The packet is then sent entirely to the Traffic Manager, which positions the entire structure in a queue associated with an output port. After the packet is serviced by the scheduler, it is separated again so that the Headers can be processed by the Egress block. As with the Ingress buffer, the payload remains in the Egress buffer unchanged. After the Header passes through the necessary stages in the Egress block, the packet is then reassembled to be forwarded or marked for drop, if applicable.

As already mentioned, the most important data for the AQM is the queuing delay metadata (or queue depth) which is only available at the Egress block. The way the AQM works is by setting a FLAG which informs to the Egress block that the packet must be dropped. This action is performed only after the end of Header processing in the Egress block, that is, the buffer resources (memory) that are being used by the Payload are finally released. In this work, we define this waste of resources as \textbf{The Egress drop problem}.

%However, when deciding to discard a packet in the Egress block, AQM must indicate this decision (activating a FLAG) so that the action of making the packet “disappear” is carried out. 

%is generally the main information used as input to the AQM algorithm to decide whether the packet should be dropped or not. On programmable devices, this information is captured by the Traffic Manager and made available only in the Egress block.}

%Knowing the entire path taken by the packet in the pipeline of a programmable switch, it is clear to understand why AQM mechanisms are commonly implemented in the Egress block. The queuing delay (or queue depth) is generally the main information used as input to the AQM algorithm to decide whether the packet should be dropped or not. On programmable devices, this information is captured by the Traffic Manager and made available only in the Egress block.

%Removido. In this case, it is clear that each packet dropped by an AQM in the Egress block occupied resources (CPU, memory, and bandwidth) in the Ingress and Egress. 

Understanding the causes and effects that this problem brings, we argue that it is possible to improve the use of shared resources (switch pipeline). The idea we defend is that the \textbf{decision} of the drop must be separated (in different blocks) from the \textbf{action} of the actual drop, thus having a disaggregated concept of AQM. The materialization of this new concept is described in Section \ref{sec:iRED}, in which we present the iRED algorithm.

\subsection{L4S architecture} \label{sec:l4s}

\begin{figure}[!htpb]
    \centering
    \includegraphics[width=.8\columnwidth]{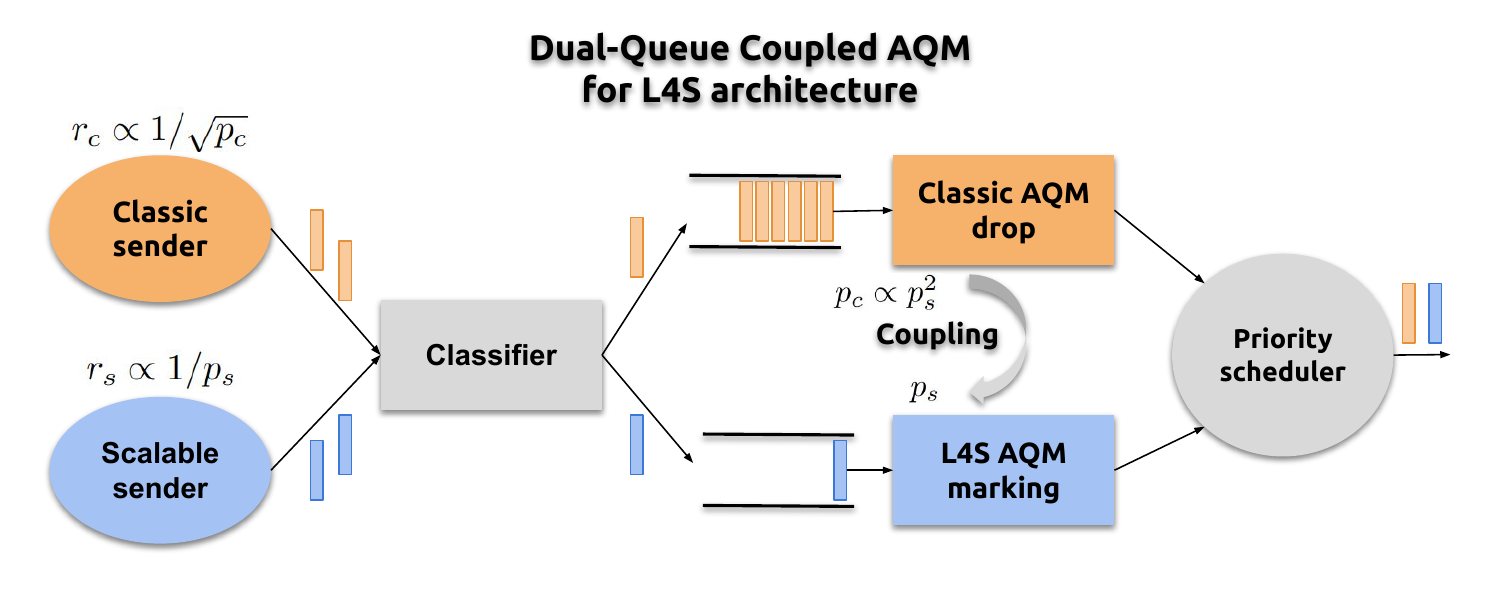}
    \caption{Dual-Queue AQM in L4S architecture. Adapted from \cite{rfc9332}.}
    \label{fig:dualqaqm}
\end{figure}

As briefly mentioned previously, the L4S architecture (shown in  Fig. \ref{fig:dualqaqm}) introduces incremental changes to both the hosts' CC algorithm and the AQM at the network nodes. The modifications proposed by L4S were motivated by some requirements, such as L4S-ECN packet identification, accurate ECN feedback, fall-back to Reno-friendly on Loss, fall-back to Reno-friendly on classic ECN bottleneck, reduce RTT dependence, scale down to the fractional window and detecting loss in units of time \cite{prague:2018}.

In this context, L4S introduces two distributed mechanisms that work together to achieve the requirements listed above. The first of these reside in the host scope, being the scalable CC algorithm, TCP Prague\footnote{The name is after an ad hoc meeting of IETF in Prague in July 2015.}\cite{prague:2018}. The TCP Prague is a modified version of Data Center TCP (DCTCP) \cite{dctcp:2010} for safe use over the Internet. As is well-known (by TCP researchers) DCTCP is suitable only for data centers, where the administrator can arrange the network to work properly for frequent ECN-marking. However, this is not so simple for the public Internet, as DCTCP flows would certainly starve classical flows. For this reason, TCP Prague presents minor modifications from DCTCP to meet the requirements listed above.

The second resides in the network nodes as a Dual-Queue coupled AQM \cite{rfc9332}, that is responsible for maintaining a harmonious coexistence between the flavors of CC, Classic and Scalable. The Dual Queue coupled AQM mechanism, specified in the RFC9332 \cite{rfc9332}, was designed to solve the coexistence problem, accommodating flows into separated queues for Classic (larger queueing delay) and Scalable (small queueing delay) CC flavors, as can be seen in Fig. \ref{fig:dualqaqm}.

Despite the use of distinct queues with varying depths (shallow and deeper), bandwidth consumption remains uniform across flows. Achieving equitable resource allocation, or harmonious coexistence, involves the interplay between the Classic and Scalable queues. This interaction enables the Classic queue to perceive the square of congestion levels in the Scalable queue. This squared is then offset by the sending rate of the classic sender ($r_c$) in response to a congestion signal, characterized by $r_c \propto 1/\sqrt{p_c}$, where $p_c$ denotes the loss level of the Classic flow. On the other hand, the Scalable sender rate ($r_s$) follows an inverse linear approach, characterized by $r_s \propto 1/p_s$, where $p_s$ denotes the loss level of Scalable flow. It is this linearity that characterizes scalability in response to congestion.

\section{iRED - Ingress Random Early Detection} \label{sec:iRED}

iRED was designed under three fundamental premises: $i)$ Perform probabilistic packet dropping with minimal overhead; $ii)$ Support and adhere to current Internet congestion control mechanisms, such as the L4S framework; $iii)$ Be fully implemented in the data plane hardware. Based on these guiding requirements, this section describes the details and challenges of implementing iRED on the Tofino2 programmable switch \footnote{The previous version of iRED \cite{iRED:2022} was deployed in a software switch environment.}.

\begin{figure}[!htpb]
    \centering
    \includegraphics[width=1\columnwidth]{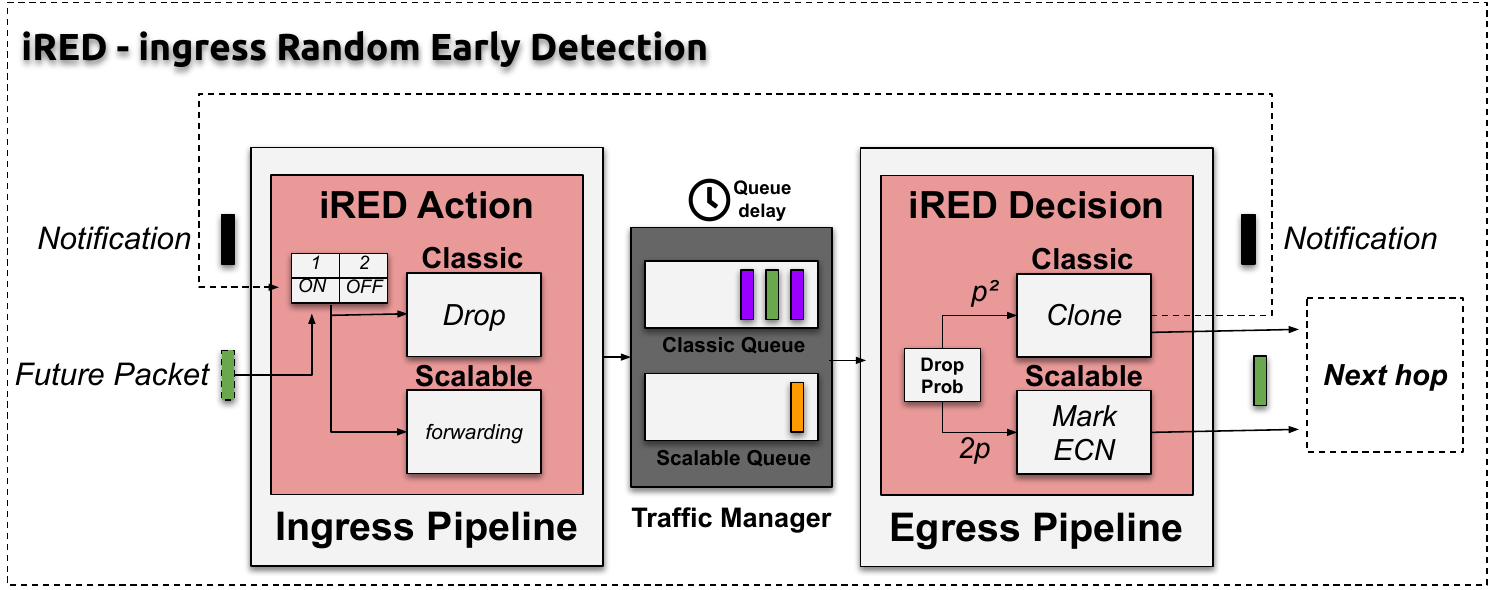}
    \caption{iRED design. Disaggregating the action of a drop decision reduces wasted resources.}
    \label{fig:iRED}
\end{figure}

Regarding the first premise, we understand that to minimize overhead on the switch, iRED should be able to discard packets as soon as possible. Leveraging the programmable switch pipeline, we believe that the most suitable place to perform the drop action is in the Ingress block. However, the data (queue metadata) necessary to calculate the drop probability is available after the Traffic Manager, that is, in the Egress block. In this context, we decided to divide iRED's operation into two parts, making it a disaggregated AQM. As can be seen in Fig. \ref{fig:iRED}, \textbf{decisions are made at the Egress, while actions are performed at the Ingress.}

%iRED, previously implemented and deployed in software switch environment \cite{iRED:2022}, is the first RED-like algorithm P4-based that implements the concept of disaggregated AQM in PDP hardware. Disaggregated in this work means performing the AQM operations in different blocks of the PDP architecture. As can be seen in Fig. \ref{fig:iRED}, \textbf{decisions are made at the Egress, while actions are performed at   the Ingress}. 

In alignment with the second premise, we implemented the AQM requirements presented previously in Sec. \ref{sec:l4s} to provide support for L4S. First of all, the classification process is performed in the Ingress block, in which the logic identifies the type of flow and enqueues it to the corresponding output queue. Furthermore, the coupling mechanism is implemented in the Egress block. In this scenario, iRED dynamically adjusts the drop probability or marking based on the flow type (Classic or Scalable).

%have incorporated specific mechanisms within iRED. Initially, a traffic classification process (distinguishing between classic and scalable) is conducted through a recognition mechanism located in the Ingress block. Secondly, we have implemented distinct queues with different priorities for classic and scalable traffic. Lastly, we employ an ECN bit marking mechanism situated in the Egress block.

Finally, iRED is fully implemented in the hardware improving the autonomy in performing AQM functions solely within the data plane, thereby eliminating the control plane or external mechanisms to make specific tasks. In this context, it is well-established that AQM logic requires the utilization of intricate mathematical operations, including multiplications, divisions, and square roots. Furthermore, certain sections of the logic require the implementation of more sophisticated functions, such as exponential moving averages or similar calculations. We overcome the challenges imposed by the architecture and implement iRED entirely in the data plane using available resources, such as bitshift to represent mathematical operations and compute the Exponentially Weighted Mean Average (EWMA).

%we encountered several challenges within the TNA architecture, including constraints on mathematical operations, limitations on memory usage, restrictions on floating-point operations, among others. However, we were able to overcome the challenges imposed by the architecture and implement iRED entirely in the data plane using available resources, such as bitshift to represent mathematical operations and recirculation to communicate the Ingress and Egress blocks.}

%In this paper, we redesigned iRED to be fully compliant with the L4S framework (Low Latency, Low Loss, and Scalable Throughput) \cite{L4S:2023}, being able to classify traffic between Classic (drop) and Scalable\footnote{TCP Prague version.} (mark ECN bit), reaching fairness among different flows by a coupled packet dropping/marking mechanism. In the iRED implementation, the coupling is arranged by the output probabilities in the Egress block.

For a more comprehensive understanding, we will initiate the description of iRED's operation from the Egress block, specifically commencing with the drop or mark decision (decision module). At the Egress, iRED computes the Exponentially Weighted Mean Average (EWMA) of the queue delay (or queue depth\footnote{The programmer can choose whether to use iRED's delay-based or depth-based approach.}) for each individual packet, entirely within the data plane. The inherent absence of division and floating-point operations poses challenges in calculating average values within the data plane. To surmount this limitation, as applied in \cite{bussegrawitz2022pforest}, we employ an approximation method following Eq. \ref{eq:ewma}:

\begin{equation}
    \centering
    S_t = \alpha \cdot Y_t + (1 - \alpha)\cdot S_{t-1}
    \label{eq:ewma}
\end{equation}

where $S_t$ is the updated average queue delay, $S_{t-1}$ is the previous average queue delay and $Y_t$ is the current queue delay. The constant $\alpha \in [0,1]$ determines how much the current value influences the average. We use $\alpha=0.5$, such multiplication can be replaced by bit shifts operations. The output of the EWMA will represent the average queue delay over time. If the value observed (average queue delay) is between a set of min-max thresholds defined, iRED will compute the drop probability according to the RED approach and will based on the coupling mechanism generate different congestion signal intensities (drop or marking).

Once the iRED decision module (Egress) has detected that a packet must be dropped (Classic), iRED must notify the action module (Ingress) to perform this action. The first challenge in the PDP context is to achieve communication between the Ingress and Egress blocks with minimum overhead. Obviously, iRED will not drop the packet that generated the discard decision, but a future packet \cite{conquest:2019}. Discarding future packets is one of the main features differentiating iRED from other state-of-the-art AQMs. For the congestion notification to reach the Ingress block, iRED creates a congestion notification packet (clone packet with only 48 bytes) and sends it through an internal recirculation port to reach the Ingress block. 

\begin{algorithm}[ht]
\small
\label{alg:Egress}
\caption{\textsc{DECISION TO DROP - EGRESS}}
\begin{algorithmic}[1]
    \State $minThsld = TARGET\_DELAY | QUEUE\_DEPTH$
    \State $maxThsld = 2*minThsld$ 
    \State $dropProb = 0$
    
    \For {$each$ $pkt$} 
        \If{$pkt == pktCloned$}
            \State $recirculate(pkt)$
        \Else
            \State $EWMA = 0.5 \cdot queue\_delay + (1 - 0.5)\cdot EWMA_{t-1}$
        
            \If{($EWMA \geq minThsld$) \textbf{and} ($EWMA \leq maxThsld$)}
                
                \State $randClassic = random(0,65535)$
                \State $randL4S = randClassic/2$
                
                \If{L4S}
                    \If{$randL4S < dropProb$}
                        \State $markECN(pkt)$
                        \State $dropProb = dropProb - 1$
                    \Else
                        \State $dropProb = dropProb + 1$
                    \EndIf
                \Else
                    \If{$randClassic < dropProb$}
                        \State $dropProb = dropProb - 1$
                        \State $clone(pkt)$
                    \Else   
                        \State $dropProb = dropProb + 1$
                    \EndIf
                \EndIf
            \EndIf                
            
            \If{$EWMA > maxThsld$}
        
                \If{L4S}
                    \State $markECN(pkt)$
                \Else
                    \State $clone(pkt)$
                \EndIf
            \EndIf
        \EndIf
    \EndFor
\end{algorithmic}
\end{algorithm}

Algorithm 1 presents the iRED decision module, which operates within the Egress block. This module continuously monitors the queue delay (or depth) and maintains an updating register that stores the probability for dropping (Classic) or marking the ECN bit (Scalable).

Algorithm 1 functions as follows: when the packet is identified as a clone, it is recirculated to the Ingress block (lines 5-6). This action signifies that forthcoming packets should be dropped in the Ingress for the designated output port, thereby consuming only 48 bytes per packet. For regular packets, not cloned, the current queue delay is employed to calculate the EWMA based on Equation \ref{eq:ewma} (line 8). If the EWMA value falls within the defined minimum and maximum thresholds (line 9), iRED proceeds to calculate the probability of dropping or marking with ECN. The decision module employs a random number generator to compute distinct probabilities for each traffic type (lines 10-11). It is noteworthy to clarify that for L4S packets, the marking probability is twice as high as that for classic packets (coupling mechanism). Consequently, the random number used in the computation of the L4S marking probability is half of the random number employed for determining the drop probability, as stipulated by the L4S framework \cite{L4S:2023}.

The subsequent step in the algorithm involves identifying the packet type, which could be L4S or Classic. If the packet type is determined to be L4S (line 12), the decision module proceeds to compare the randomly generated L4S number with the drop probability value stored in a register (line 13). If this comparison yields a true result, indicating that the L4S packet should be marked, the ECN bit of the L4S packet is set to 1 (line 14), and the drop probability value stored in the register is decremented by one unit (line 15). Conversely, if the condition is false (line 16), the drop probability value in the register is incremented by one unit (line 17).

For Classic traffic, the logic is analogous (lines 20-24). However, instead of marking the ECN bit, the decision module executes a clone operation (line 22). In the clone operation, the original packet remains unaltered and proceeds to be forwarded as usual to its final destination. Simultaneously, the clone packet is modified to carry only notification information destined for the action module.

In cases where the EWMA value exceeds the established maximum threshold (line 28), a uniform action is taken: either all packets are marked as L4S or all packets are cloned as Classic, depending on the traffic type (lines 29-32).

The action module, situated in the Ingress block, maintains the congestion state table on a per-port/queue basis and activates the drop flag (ON) for the corresponding port/queue. The current packet is forwarded to the next hop without introducing any additional delay. Subsequently, future packets intended for the same output port/queue, where the drop flag is set to ON, will be dropped (classic), and the drop flag will be reset to OFF. This mechanism, facilitated by iRED, ensures that the Ingress pipeline can proactively mitigate imminent queue congestion.

\begin{algorithm}[ht]
\small
\label{alg:Ingress}
\caption{\textsc{Action to drop - Ingress}}
\textbf{Input:} \textit{pkt, pktRecirc}
\begin{algorithmic}[1]
    \For {$each$ $pkt$} 
    
        \If{$pkt == pktRecirc$}
            \State $dropFlag[output\_port] = 1$ \Comment{Flag to drop ON}
            \State $drop(pktRecirc)$
        \EndIf
    \State $ip\_forward$ 
        \State $dropPort = dropFlag[output\_port]$
        \If{$dropPort == 1$}

            \If{L4S}
                \State $forwarding$
            \Else
                \State $drop(pkt)$ \Comment{Packet dropped}
                \State $dropFlag[output\_port] = 0$ \Comment{Flag to drop OFF}
            \EndIf
        \EndIf    
    \EndFor
\end{algorithmic}
\end{algorithm} 

Now, we will explain the action part of iRED (listed in Algorithm 2), which runs at the Ingress block. The initial step in Algorithm 2 involves verifying whether the incoming packet has been recirculated from the Egress block (line 2). We employ a register with a length matching the number of ports, where each port is associated with an index. If the packet has been recirculated, the drop flag is activated by setting the corresponding value in the index register to 1 (line 3). Following this, the recirculated packet serves its purpose and is subsequently discarded (line 4).

The remainder of Algorithm 2 primarily focuses on the routine forwarding of packets (line 6), where the output port is determined. In this step, the algorithm performs an evaluation to ascertain the status of the drop flag associated with the specified output port. Should the flag be in the activated state (indicated by a value of 1), the packet undergoes a dropping procedure (as delineated in lines 12-13), and concurrently, the register is restored to its initial state. It is important to note that only one packet is dropped at a time, and subsequent packets destined for the same output port will only be dropped if a recirculated packet is detected in the Ingress pipeline, signaling congestion.

For Scalable flows, iRED does not drop packets, as expected; instead it forwards the packet to the scalable queue. In summary, iRED is the only current AQM P4-based that drops packets in the Ingress block, fully deployable in the programmable data plane hardware and is L4S-capable.

\section{Related Work} \label{sec:AQMOperation}

%2016: PI2 (v1model)
%2018: P4-CoDel (v1model)
%2019: P4-ABC (TNA)
%2020: 
%2021: PV-AQM (TNA), P4-CoDel (TNA)
%2022: (dual)PI2 (TNA), iRED (v1model)
%2023: CoDel++ (v1model), iRED(TNA2)

\begin{figure}[ht]
    \centering
    \includegraphics[width=1\columnwidth, frame]{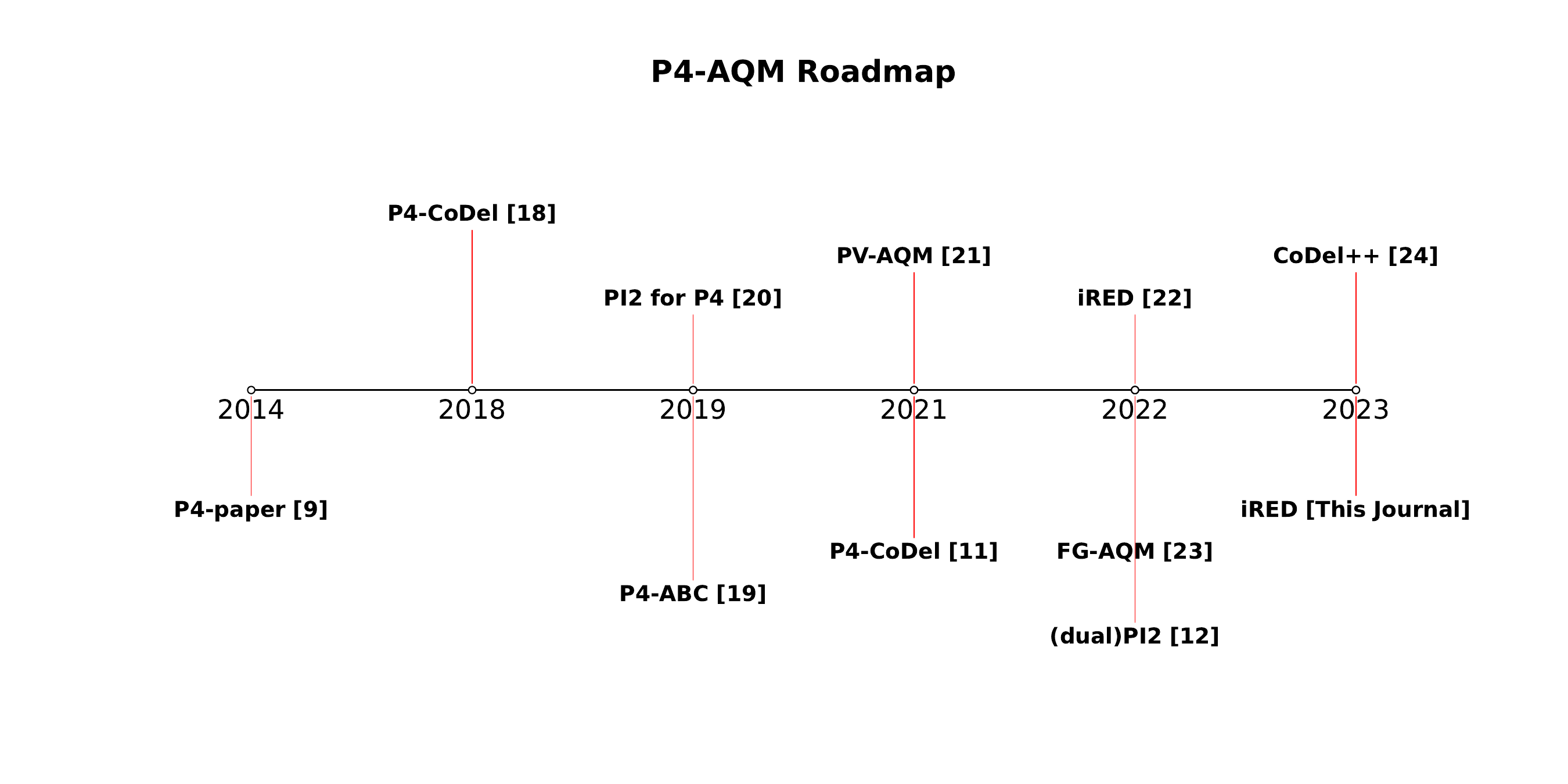}
    \caption{P4-AQM Roadmap. Timeline of the scientific community's main efforts in AQMs programmable data plane.}
    \label{fig:p4-aqm-roadmap}
\end{figure}

In this section, we give a brief overview of the state-of-the-art regarding P4-based AQM algorithms. We will present the main characteristics of the probabilistic drop operation for each approach, as well as point out the challenges and weaknesses that still exist in these proposals. At the end of the section, we present a comparative table between the state-of-the-art approaches and iRED. We will follow a chronological path to discuss the evaluation of the approaches, as can be seen in Fig. \ref{fig:p4-aqm-roadmap}.

Since the paper introducing P4 in 2014 \cite{P4-paper}, the scholarly community has exhibited substantial engagement with the subject of AQM within the programmable data plane. The initial noteworthy endeavor was published in 2018 \cite{p4-CoDel:2018}. The work titled ``\textit{P4-CoDel: Active Queue Management in Programmable Data Planes}" presented a P4 implementation of the CoDel \cite{CODEL:2012} algorithm in a software switch environment (v1model architecture).

In the following year, 2019, two other works \cite{p4-abc:2019, pi24p4:2019} emerged in the same context. The first, published in July 2019, whose title is ``\textit{Implementation and Evaluation of Activity-Based Congestion Management Using P4 (P4-ABC)}'', presents an AQM prototype in P4 based on the ABC strategy using software switch (v1model architecture) and some externs (out of data plane domain) to workaround some data plane limitations. In December 2019, the work titled ``\textit{PI2 for P4: An Active Queue Management Scheme for Programmable Data Planes}" was presented as the first implementation of the  Proportional Integral (PI) controller in P4 in a software switch environment (v1model architecture). In this version, part of the AQM logic was implemented in the control plane due to the data plane constraints in the math operations.

It was only in 2021 that a more robust performance analysis of a hardware version of P4-CoDel (using the TNA architecture) was presented \cite{kundel2021p4codel} with the title ``\textit{P4-CoDel: Experiences on Programmable Data Plane Hardware}". Still in 2021, the thesis ``\textit{Making a Packet-value Based AQM on a Programmable Switch for Resource-sharing and Low Latency}" was defended proposing PV-AQM \cite{pvaqm:2021}, an AQM based on Per-Packet Value (PPV) and Proportional Integral Controller Enhanced (PIE) in programmable data plane hardware (TNA). As in \cite{pi24p4:2019}, this work also used the control plane to calculate the drop probability with the PIE controller.

In August 2022, a more in-depth evaluation of PI2 in hardware (TNA) has been published \cite{DualPI2} in the work ``\textit{Active Queue Management on the Tofino programmable switch: The (Dual)PI2 case}". Furthermore, an extension of PI2 (dualPI2) to support the L4S framework was also developed and presented in the same work. In both versions, the control plane continues to be used to assist in computing complex mathematical operations performed by the PI controller. In the same year, the first version of iRED was presented in a software switch environment (v1model) \cite{iRED:2022}, evaluating an adaptive video streaming scenario in the work ``\textit{iRED: Improving the DASH QoS by dropping packets in programmable data planes}". In this previous version, iRED obtained superior results in relation to the state-of-the-art AQMs and the Tail Drop approach. Still in 2022, FG-AQM was published \cite{fg-aqm:2022} in the work \textit{``Fine-Grained Active Queue Management in the Data Plane with P4"}. In this case, FG-AQM uses a PI controller to compute the drop probability in a software switch environment (v1model).

In 2023, other works similar to this one are being discussed and presented in the scientific community. We highlight CoDel++ \cite{CoDel++:2023}, a new version that combines the use of priority queues in the CoDel algorithm in hardware (TNA) that was proposed in the work ``\textit{Interplay Between Priority Queues and Controlled Delay in Programmable Data Planes}".

In light of all these works, we tried to observe which of them have publicly available versions of their P4 code for programmable Tofino switches. At the time of this writing, the only ones are P4-CoDel and PI2. For this reason, in this section, we look into more details at the internal working mechanisms of these approaches so that we can compare with iRED.

\subsection{P4-CoDel}
%2018(v1model) 2020(TNA)
Controlled Delay (CoDel) is an AQM specified by the IETF in RFC 8289, that uses the \textit{sojourn time} \cite{CODEL:2012} and sliding window measurements to control the congestion. \textit{Sojourn time} is given by the time that any packet waits in the queue, the queue delay. CoDel, therefore, measures the \textit{sojourn time} and tracks whether it is consistently\footnote{Lasting longer than a typical RTT.} sitting above some tiny target \cite{TCPCC}. As TCP throughput depends inversely on the square root of the loss rate \cite{rfc5348}, CoDel steadily increases its drop rate in proportion to the square root of the number of drops since the target was exceeded \cite{TCPCC}.

\begin{figure}[ht]
    \centering
    \includegraphics[width=.7\columnwidth]{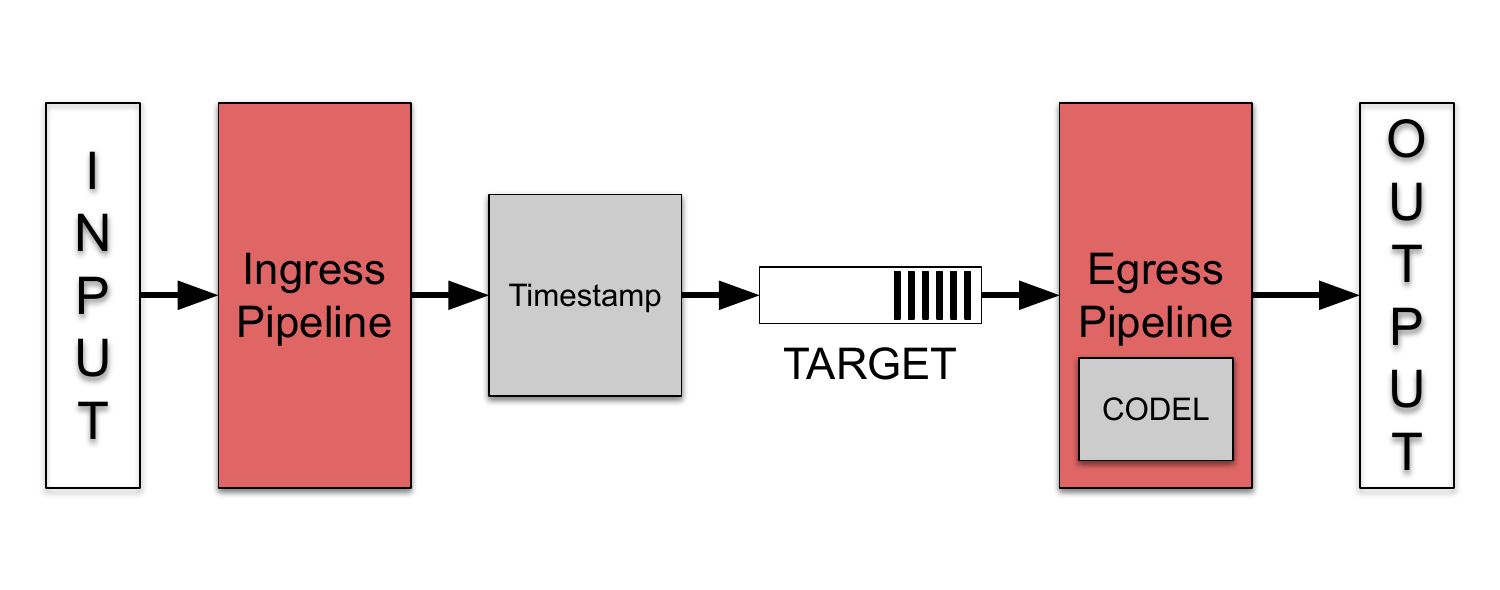}
    \caption{P4-CoDel overview. P4-CoDel is an implementation of the CoDel algorithm in a Tofino Switch, positioned at the Egress block.}
    \label{fig:p4-codel}
\end{figure}

P4-CoDel is the implementation of CoDel in TNA \cite{kundel2021p4codel} as can be seen in Fig. \ref{fig:p4-codel}, trying to keep the queueing delay below a specified (TARGET parameter) in a one-time interval (INTERVAL parameter), following these steps for each packet: 

\begin{enumerate}
    \item If the queueing delay is below the threshold, a packet is never dropped;
    \item If the threshold is reached by more than a certain interval time unit, the first packet will be dropped;
    \item From then on, the interval between dropping packets gets smaller until the threshold delay is reached.
\end{enumerate}

However, there is no free lunch. To make all computations in the data plane, P4-CoDel uses a math unit within the stateful ALU to approximate the square root function in multiple match-action stages. Moreover, the authors employed the P4 longest prefix match table feature to map approximations for the square root.

\subsection{PI2}
%2016(v1model) 2022(TNA)
PI2 is a linearized AQM for both classic (TCP Cubic) and scalable TCP (TCP Prague), based on the Proportional Integral (PI) controller \cite{pi2:2016}. The PI2 uses queueing information (delay) per packet periodically (T interval) in conjunction with PI gain factors ($\alpha$ and $\beta$) to trigger the packet drop policy, as described in Equation \ref{eq:pi}.

\begin{equation}
    \centering
    p = p+\beta(\tau_{t-1}-\tau_t) + \alpha(\tau_0 - \tau_1) 
    \label{eq:pi}
\end{equation}

Any alteration in the queue, be it an increase or decrease, is promptly rectified through the application of a proportional gain factor denoted as $\beta$, while any persisting deviation from the target, referred to as residual error, is gradually attenuated towards the target through the utilization of an integral gain factor denoted as $\alpha$ \cite{DualPI2}.  The output probability of the basic PI controller is squared when dropping classic TCP packets or doubled when marking scalable TCP traffic. PI2 AQM proved that by simply squaring the output PI probability, the PIE auto-tuning and several heuristics could be removed.

\begin{figure}[ht]
    \centering
    \includegraphics[width=.7\columnwidth]{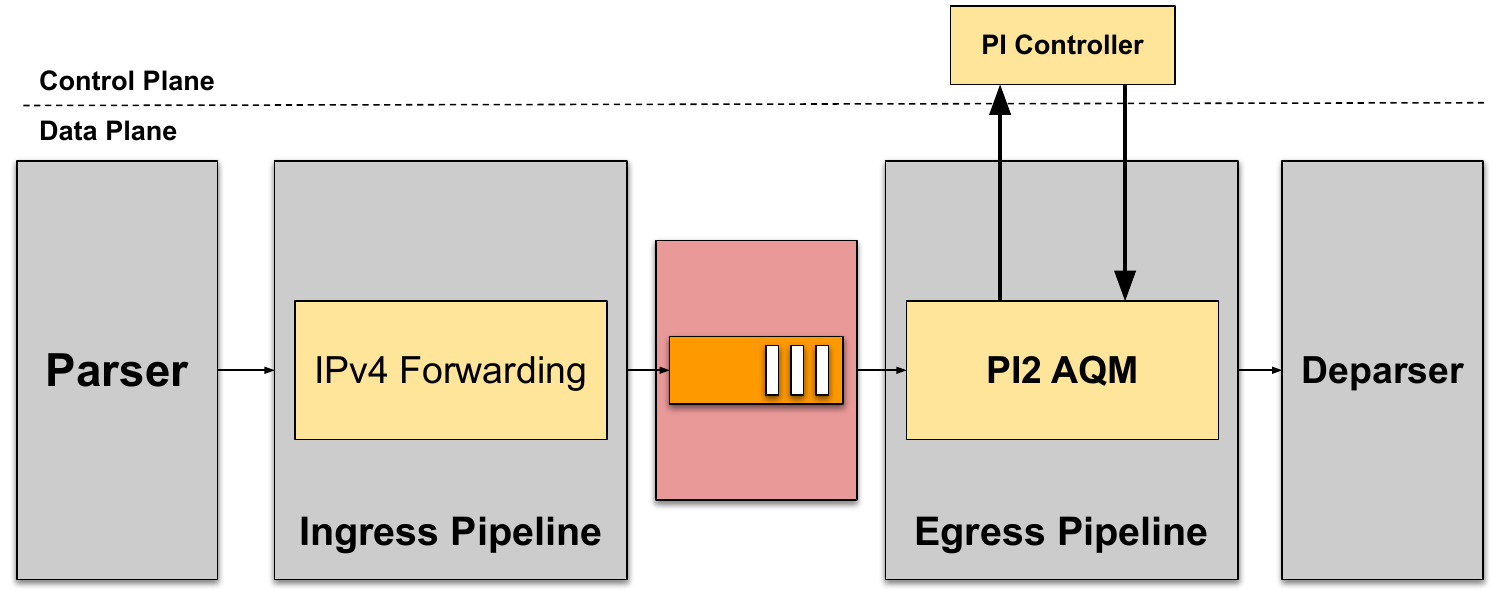}
    \caption{PI2 overview. The PI controller, positioned at the Control Plane retrieves the queue delay from the data plane to compute the drop probability.}
    \label{fig:pi2}
\end{figure}

PI2 for P4 is an implementation for TNA \cite{DualPI2} has part of the logic implemented in a control plane, as detailed in Fig. \ref{fig:pi2}, to perform the required complex arithmetic operations that can not be handled by the data plane due to the restricted set of math operations in the PDP architectures. The control plane periodically retrieves the queuing delay from the data plane and uses it in the PI Controller to determine the probability of drop (Classic) or mark \cite{L4S:2023} (Scalable).

As observed in the P4-CoDel case, the direct execution of intricate mathematical operations within the programmable data plane remains a formidable task. As elucidated by the authors in \cite{DualPI2}, these inherent constraints necessitated the utilization of the control plane for the implementation of the PI controller. Consequently, the principal limitation of PI2 manifests itself in its reliance on the control plane, thereby incurring an additional delay in the computation of the PI controller.

\subsection{A summary of AQMs}

Following an examination of the state-of-the-art AQM mechanisms implemented in programmable hardware using P4 and a comprehensive review of the accompanying source code, we have identified and emphasized key attributes. Table \ref{AQMcomparison} shows the distinguishing features among the analyzed approaches within the scope of this work.

\begin{table}[ht]
\caption{AQM comparative analysis.}
\begin{tabular}{|c|c|c|c|c|c|}
\hline
\textbf{AQM} & \textbf{Delay-based} & \textbf{Depth-based} & \textbf{Ingress Drop} & \textbf{L4S compliance} & \textbf{Fully imp. in the data plane} \\ \hline
P4-CoDel   & \checkmark   &  \textit{x} &  \textit{x} & \textit{x} & \checkmark\\ \hline
PI2 for P4     & \checkmark   &  \textit{x}  &  \textit{x}  & \checkmark & \textit{x} \\ \hline
iRED    & \checkmark  &  \checkmark  &  \checkmark & \checkmark & \checkmark \\ \hline

\end{tabular}
\label{AQMcomparison}
\end{table}

Upon scrutinizing the data presented in Table \ref{AQMcomparison}, it becomes apparent that all of the examined AQM systems incorporate support for queue delay as a fundamental metric. Nevertheless, it is noteworthy that the inclusion of queue depth as a supported metric is unique to the iRED AQM. Furthermore, iRED is the only one that currently supports dropping packets in the Ingress block. On the other hand, P4-CoDel is the only AQM that does not conform to the L4S framework. Finally, it is imperative to underscore that PI2 cannot be regarded as a fully data-plane-implemented AQM, given its reliance on the control plane for supplementary computational tasks.

\section{Evaluation} \label{sec:evaluation}

In the present study, we have conducted three types of experiments. Firstly, we evaluate resource utilization within the context of existing AQM algorithms implemented in the P4, utilizing a Tofino2 programmable switch as the experimental platform. Following this, we proceed to assess the compatibility of AQM with the L4S framework, with a specific focus on gauging fairness in the concurrent operation of classic (TCP cubic) and scalable (TCP Prague) flow types. Lastly, we conduct an evaluation of the AQM algorithms within the context of an adaptive video streaming scenario, specifically focusing on DASH.

\subsection{Resource Consumption Analysis}
We fully implemented iRED for the TNA2 architecture and performed an in-depth evaluation with the state-of-the-art Egress-based AQMs (P4-CoDel and PI2 for P4), reproducing the same setup (traffic intensity) conducted in \cite{DualPI2}. All artifacts used in this section are available in the open repository for reproducibility\footnote{https://github.com/dcomp-leris.}.

We are aware that TNA2 brings a new and interesting feature called Ghost thread which allows to obtain the egress queue metadata from the Ingress pipeline. However, a few constraints still exist. First of all, the Ghost thread provides the queue length, while P4-CoDel and (dual) PI2 need queue delay. So, they need to be adapted. Second, as far as we could understand, the Ghost thread needs to somehow update the status of the queues from egress to ingress, incurring certain overhead.  Although we believe that more investigation needs to be done regarding the performance of the Ghost thread and its usage for AQMs, in this paper we also provide an implementation of iRED (named iRED+G) compliant with the Ghost thread so that we can minimally evaluate it.

\begin{figure}[!htpb]
    \centering
    \includegraphics[width=.8\columnwidth]{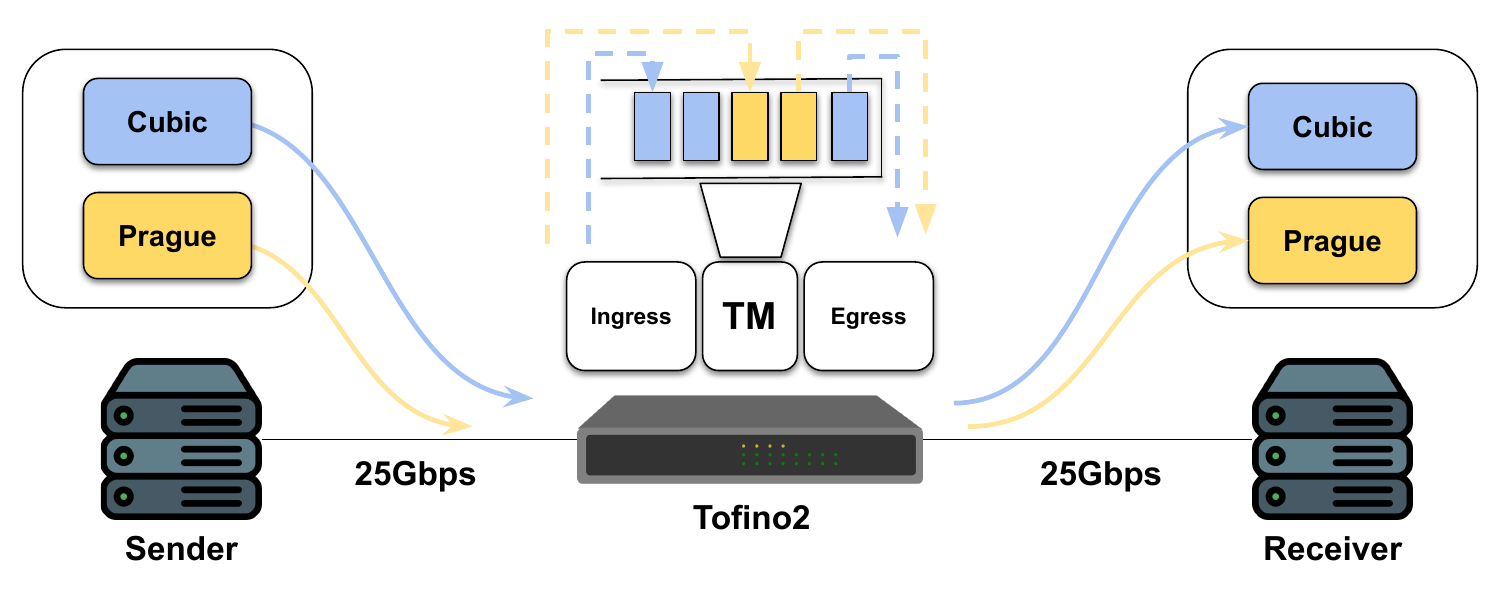}
    \caption{Evaluation setup. Cubic and Prague flows coexist in the same scenario, sharing the programmable switch bandwidth.}
    \label{fig:setup}
\end{figure}

\textbf{Environment description.} Our testbed consists of a P4 programmable switch (Edgecore DCS810 - Tofino2). The switch connects two Linux hosts, Sender and Receiver, having 25Gbps of link capacity, as shown in Fig. \ref{fig:setup}. Seeking to analyze the coexistence and fairness between different versions of TCP, each end-host sends TCP Cubic and Prague flows. We conducted our experiments over different network conditions shown in Table \ref{tab:Configurations}, varying bandwidth, RTT and MTU. The bandwidth is emulated by the P4-switch using the port shaping feature. The base RTT is emulated in the Receiver by the \textit{tc netem} tool, delaying the ACKs of TCP flows. The MTU is emulated in the end-hosts (Sender and Receiver) by the \textit{ifconfig} tool. The traffic is generated by the \textit{iperf} tool.

\begin{table}[!htpb]
\caption{Configurations}
\centering
\begin{tabular}{ l c c c }
\toprule
 \textbf{Configuration} & \textbf{Bandwidth(Mbps)} & \textbf{RTT(ms)} & \textbf{MTU(Bytes)} \\
 \midrule
   I    & 120  & 10 & 1500  \\
   II   & 120  & 50 & 1500  \\
   III  & 1000 & 10 & 1500  \\
   IV   & 1000 & 50 & 1500  \\
    V   & 120  & 10 & 800  \\
   VI   & 120  & 50 & 800  \\
   VII  & 1000 & 10 & 800  \\
   VIII & 1000 & 50 & 800  \\
    IX  & 120  & 10 & 400  \\
    X   & 120  & 50 & 400  \\
   XI   & 1000 & 10 & 400  \\
   XII  & 1000 & 50 & 400  \\
 \bottomrule
 \end{tabular}
 \label{tab:Configurations}
\end{table}

\textbf{Load description.} The load applied to the experiment is composed of 4 phases of 120 seconds each. In each phase, new flows enter the system, that is, starting with less load and ending with a high load (bottleneck condition), as used in \cite{DualPI2}. The number of Cubic and Prague flows are shown in Table \ref{tab:Load}.

\begin{table}[!htpb]
\caption{Load parameters}
\centering
\begin{tabular}{cccc}
\toprule
\textbf{Phase} & \textbf{Relative time} & \textbf{Cubic} \textbf{Flows} & \textbf{Prague Flows} \\ \midrule
1              & 0                      & 1                    & 1                     \\ 
2              & 120                    & 2                    & 2                     \\ 
3              & 240                    & 10                   & 10                    \\ 
4              & 360                    & 25                   & 25                    \\
\bottomrule
\end{tabular}
\label{tab:Load}
\end{table}

\textbf{AQMs settings.} We use a base TARGET DELAY of 20ms for all AQMs. For iRED, we set the minimum and maximum thresholds for queue delay, configuring 20 (TARGET delay) and 40 ms respectively, following the rule of thumb to set the maximum threshold as at least twice the minimum \cite{RED:93}. For PI2, we set the TARGET delay (20ms), INTERVAL (15ms), $\alpha$ (0.3125) and $\beta$ (3.125), following the
parameters used in \cite{DualPI2}.  In P4-CoDel, we set the TARGET delay (20ms) and INTERVAL (100ms), following the values used in \cite{kundel2021p4codel}.

\textbf{Ghost Thread.} As already mentioned, Tofino2 provides a new feature that enables the observation of the queue depth at the Ingress block per packet. From the flexibility that is brought by this new feature, we created a modified iRED version (iRED+G), that obtains the Egress port queue depth at the Ingres block, and then, makes the decision and the dropping both at the Ingres block. The key difference here is that we needed to adapt the iRED to use the queue depth rather than the queue delay. 

\textbf{Metrics and Measurements.} The objective of the evaluation is to analyze the consumption of switch resources for \textit{all packets discarded by the AQM methods at the Egress block}, that is, the resources that were wasted. In this context, we evaluated four metrics: wasted memory, wasted time, wasted clock cycles (latency), and wasted weight (power consumption). The wasted memory is the sum of all memory resources used by the packets (see Table \ref{tab:drops}) until being dropped, expressed in megabyte (MB). The wasted time is the sum of all time used by the packets until being dropped, expressed in milliseconds. The wasted cycles is the number of clock cycles and weight is a metric that represents the power consumption (unit-less).

\textbf{Tables in grayscale.} All tables used to present the results are colored in grayscale, in which the range of values is between light (best value) and dark (worst value).

\begin{table}[!htpb]
    \caption{Number of dropped packets}
    \centering
    \begin{tabular}{lcccccccc}
    \toprule
        \textbf{Conf} & \textbf{iRED} & \textbf{PI2} & \textbf{CoDel} & \textbf{iRED+G} \\
        \midrule
        I & 35597 & 58417 & 35861 & 37561 \\ 
        II & 11311 & 22735 & 17947 & 11575 \\ 
        III & 9495 & 7546 & 36602 & 45725 \\ 
        IV & 4103 & 1802 & 9086 & 13266 \\ 
        V & 27826 & 38060 & 21282 & 29016 \\ 
        VI & 7625 & 19246 & 20665 & 8077 \\ 
        VII & 6141 & 4538 & 33136 & 28367 \\ 
        VIII & 7296 & 2378 & 15301 & 15612 \\ 
        IX & 18314 & 26663 & 33044 & 21841 \\ 
        X & 5455 & 11852 & 10430 & 5973 \\ 
        XI & 4684 & 2639 & 23975 & 17669 \\ 
        XII & 12080 & 1510 & 29870 & 19744 \\ 
    \bottomrule
    \end{tabular}
    \label{tab:drops}
\end{table}

\textbf{Number of dropped packets.} All evaluations were performed based on the number of dropped packets in each configuration, detailed in Table \ref{tab:drops}. The variation of the numbers refers to the drop probability (randomness) for each AQM.

\subsubsection{Wasted Memory}

In this subsection, we detail the results of wasted memory for each configuration evaluated in Table \ref{tab:WastedMemory}. In the case of Egress-based AQMs, the wasted memory is calculated by doing \textit{2 * the size of the packet} (1500 bytes in the Ingress Buffer + 1500 bytes in the Traffic Manager). For iRED, the wasted memory is computed by the sum of the length of the dropped (1500 bytes) and notification (48 bytes) packets, resulting in $1500 + 48 = 1548$ bytes. For the iRED+G, the wasted memory is only the Ingress buffer, which is 1500 bytes. We conjecture that there is some internal memory used by the Ghost mechanism to share queue depth information between the Traffic Manager and Ingress, but it's an internal feature that is not exposed to the programmer. 

\begin{table}[!htpb]
\caption{Wasted Memory (MB)}
\centering
\label{tab:WastedMemory}
\begin{tabular}{lcccccccc}
\toprule
\textbf{Conf} & \textbf{iRED} & \textbf{PI2} & \textbf{CoDel} & \textbf{iRED+G} \\
\midrule
\textbf{I} & \cellcolor[gray]{0.8107}55.09 & \cellcolor[gray]{0.5000}175.24 & \cellcolor[gray]{0.6749}107.58 & \cellcolor[gray]{0.8074}56.34 \\
\textbf{II} & \cellcolor[gray]{0.9079}17.5 & \cellcolor[gray]{0.7768}68.2 & \cellcolor[gray]{0.8139}53.84 & \cellcolor[gray]{0.9082}17.36 \\
\textbf{III} & \cellcolor[gray]{0.9151}14.69 & \cellcolor[gray]{0.8942}22.78 & \cellcolor[gray]{0.6692}109.8 & \cellcolor[gray]{0.7758}68.58 \\
\textbf{IV} & \cellcolor[gray]{0.9367}6.34 & \cellcolor[gray]{0.9391}5.4 & \cellcolor[gray]{0.8827}27.24 & \cellcolor[gray]{0.9017}19.89 \\
\textbf{V} & \cellcolor[gray]{0.8921}23.59 & \cellcolor[gray]{0.7957}60.89 & \cellcolor[gray]{0.8651}34.05 & \cellcolor[gray]{0.8931}23.21 \\
\textbf{VI} & \cellcolor[gray]{0.9364}6.46 & \cellcolor[gray]{0.8735}30.79 & \cellcolor[gray]{0.8676}33.06 & \cellcolor[gray]{0.9364}6.46 \\
\textbf{VII} & \cellcolor[gray]{0.9397}5.19 & \cellcolor[gray]{0.9343}7.26 & \cellcolor[gray]{0.8160}53.01 & \cellcolor[gray]{0.8944}22.69 \\
\textbf{VIII} & \cellcolor[gray]{0.9372}6.15 & \cellcolor[gray]{0.9433}3.8 & \cellcolor[gray]{0.8898}24.48 & \cellcolor[gray]{0.9208}12.48 \\
\textbf{IX} & \cellcolor[gray]{0.9319}8.19 & \cellcolor[gray]{0.8980}21.33 & \cellcolor[gray]{0.8848}26.43 & \cellcolor[gray]{0.9305}8.73 \\
\textbf{X} & \cellcolor[gray]{0.9468}2.44 & \cellcolor[gray]{0.9286}9.48 & \cellcolor[gray]{0.9315}8.34 & \cellcolor[gray]{0.9469}2.38 \\
\textbf{XI} & \cellcolor[gray]{0.9477}2.09 & \cellcolor[gray]{0.9476}2.11 & \cellcolor[gray]{0.9035}19.18 & \cellcolor[gray]{0.9348}7.06 \\
\textbf{XII} & \cellcolor[gray]{0.9391}5.4 & \cellcolor[gray]{0.9500}1.2 & \cellcolor[gray]{0.8913}23.89 & \cellcolor[gray]{0.9327}7.89 \\
\bottomrule
\end{tabular}
\end{table}

In general, as can be seen in Table \ref{tab:WastedMemory}, Egress-based AQMs need more memory to perform drops, given the same load. This happens because the AQM operations (decision and action) are combined in the Egress block. As packets dropped by iRED only cross the Ingress block, there is up to 10x less memory usage (Configuration VII). 

\subsubsection{Wasted Time}

In the case of the Egress-based AQM, the wasted time is defined by the queue delay computed for each discarded packet. In other words, it means the time that a given packet stayed in the output queue before being dropped. However, in TNA there is no intrinsic metadata to represent the queue delay. In this case, the traditional way \cite{kundel2021p4codel, DualPI2} to do it is to compute the difference between \textit{egress global timestamp (egTstmp)} and \textit{ingress global timestamp (igTstmp)}. This difference represents the sum of the time spent in: Ingress parser latency; Ingress processing latency; Ingress deparser latency; and Traffic Manager latency. We create an internal bridge header to carry the \textit{igTstmp} from Ingress to Egress, and when the packet reaches the Egress block, we get the \textit{egTstmp} to calculate the queue delay. 

In the Ingress-based AQMs, the discarded packets are not sent to the output queue, so \textbf{the queue delay is always zero}. However, the congestion notification needs to be carried to the Ingress block. iRED uses recirculation, so in this case, the wasted time is defined by the recirculation time for each notification packet sent from Egress to the Ingress block. Again, for the iRED+G, we were not able to compare it with the others, because it uses internal features that are not exposed to the programmer. 

\begin{figure*}[!htpb]
    \centering
     \subfigure[I]{
        \includegraphics[width=.3\columnwidth]{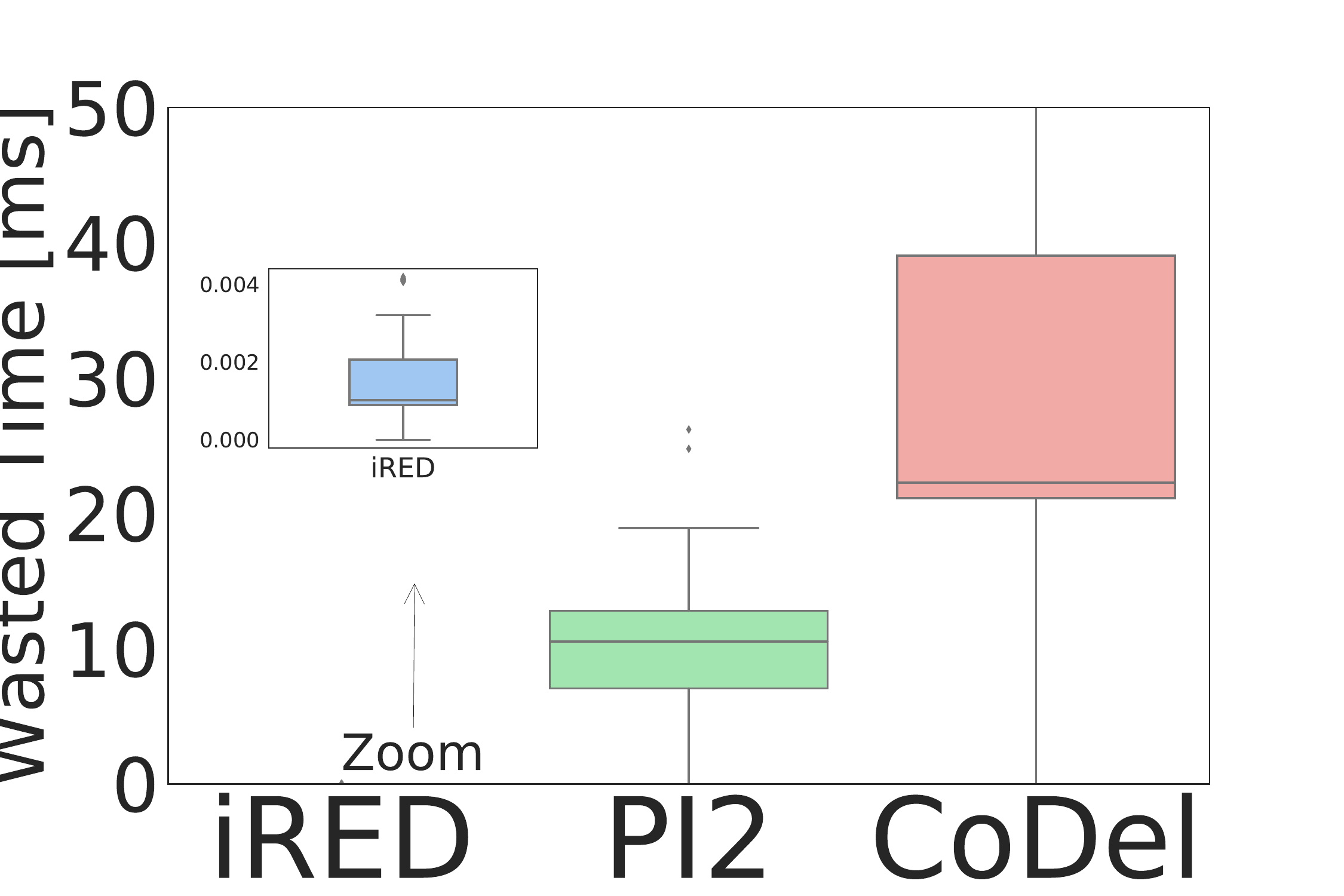}
    }
    \subfigure[II]{
        \includegraphics[width=.3\columnwidth]{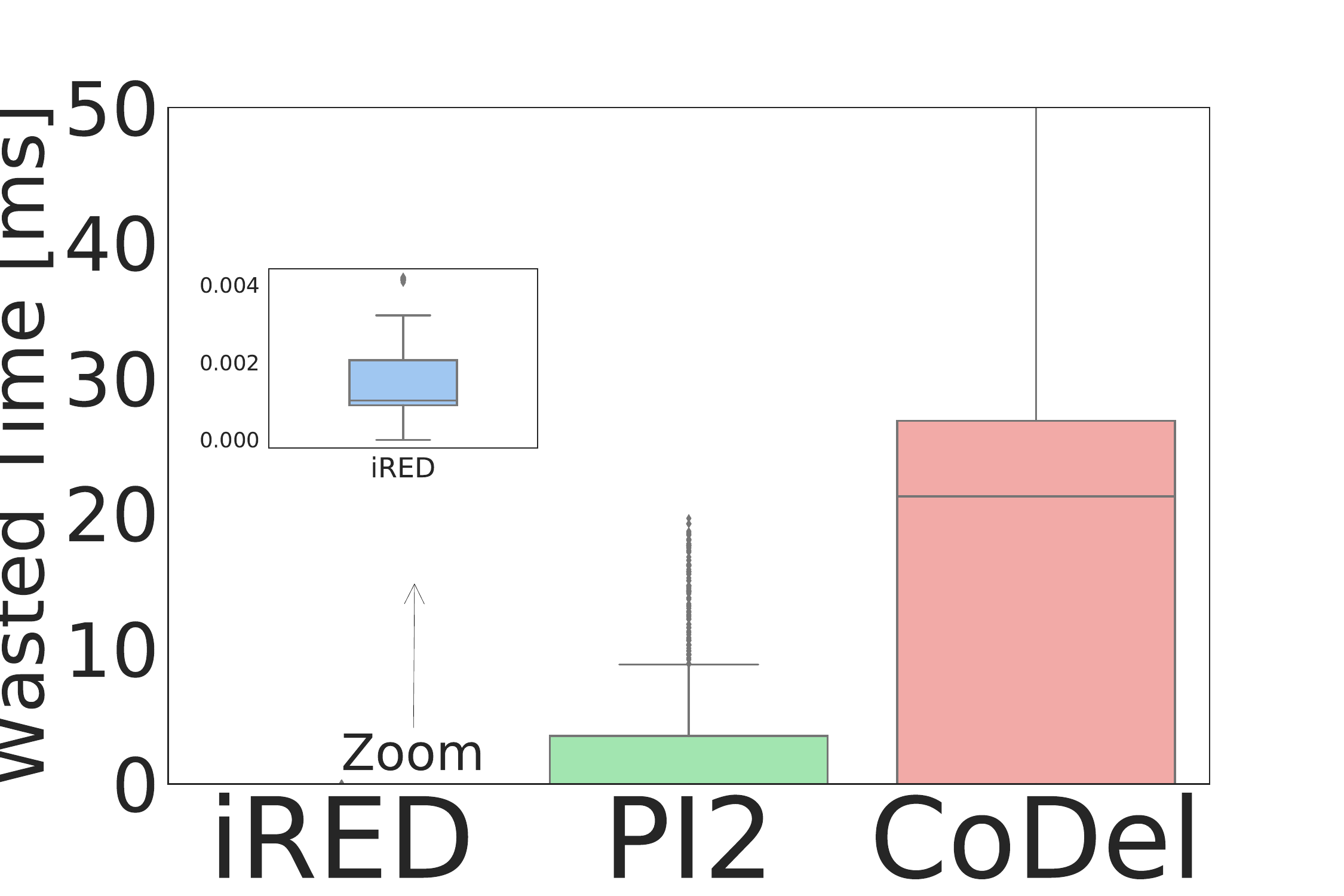}
    }
    \subfigure[III]{
        \includegraphics[width=.3\columnwidth]{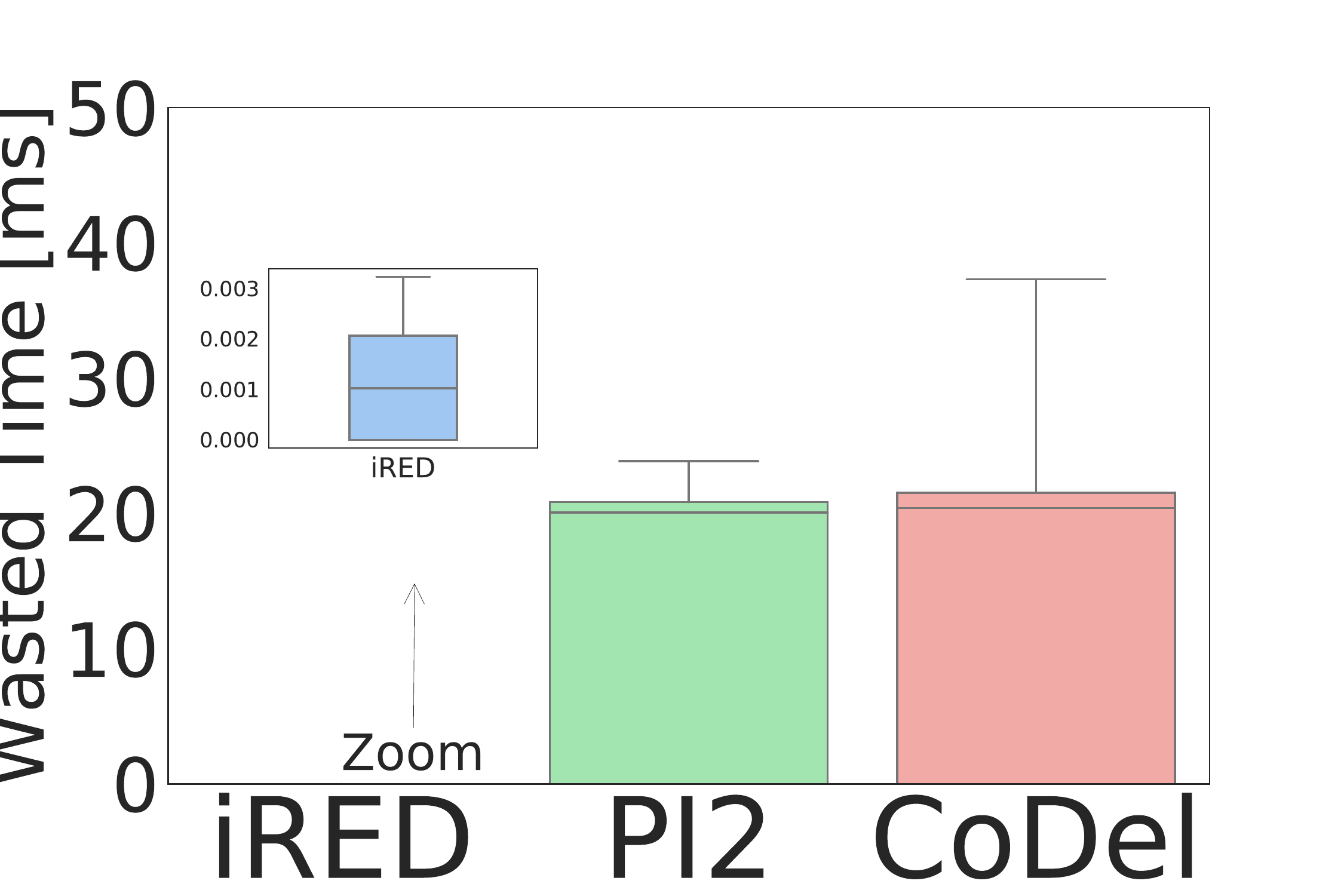}
    }
    \subfigure[IV]{
        \includegraphics[width=.3\columnwidth]{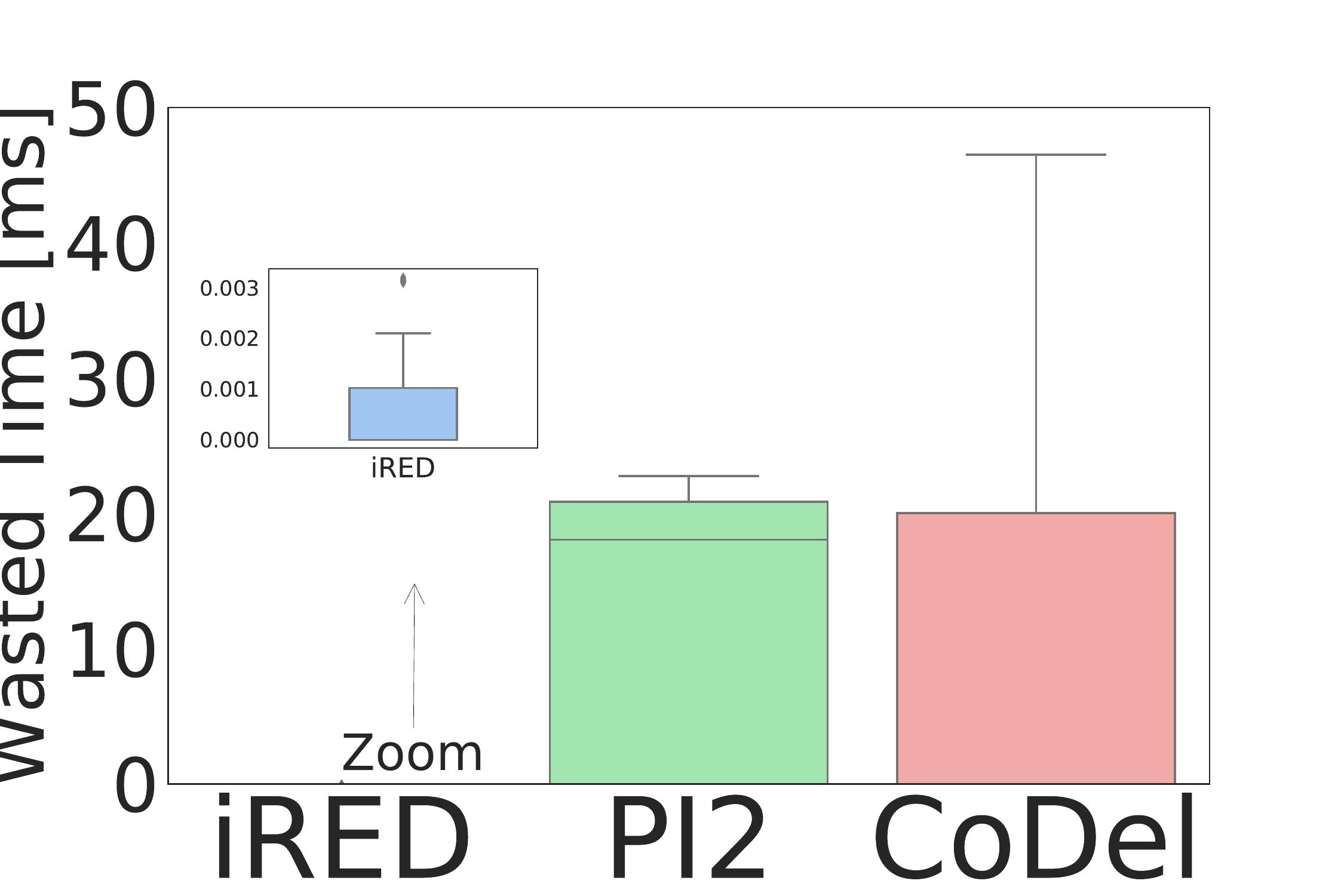}
    }
    \subfigure[V]{
        \includegraphics[width=.3\columnwidth]{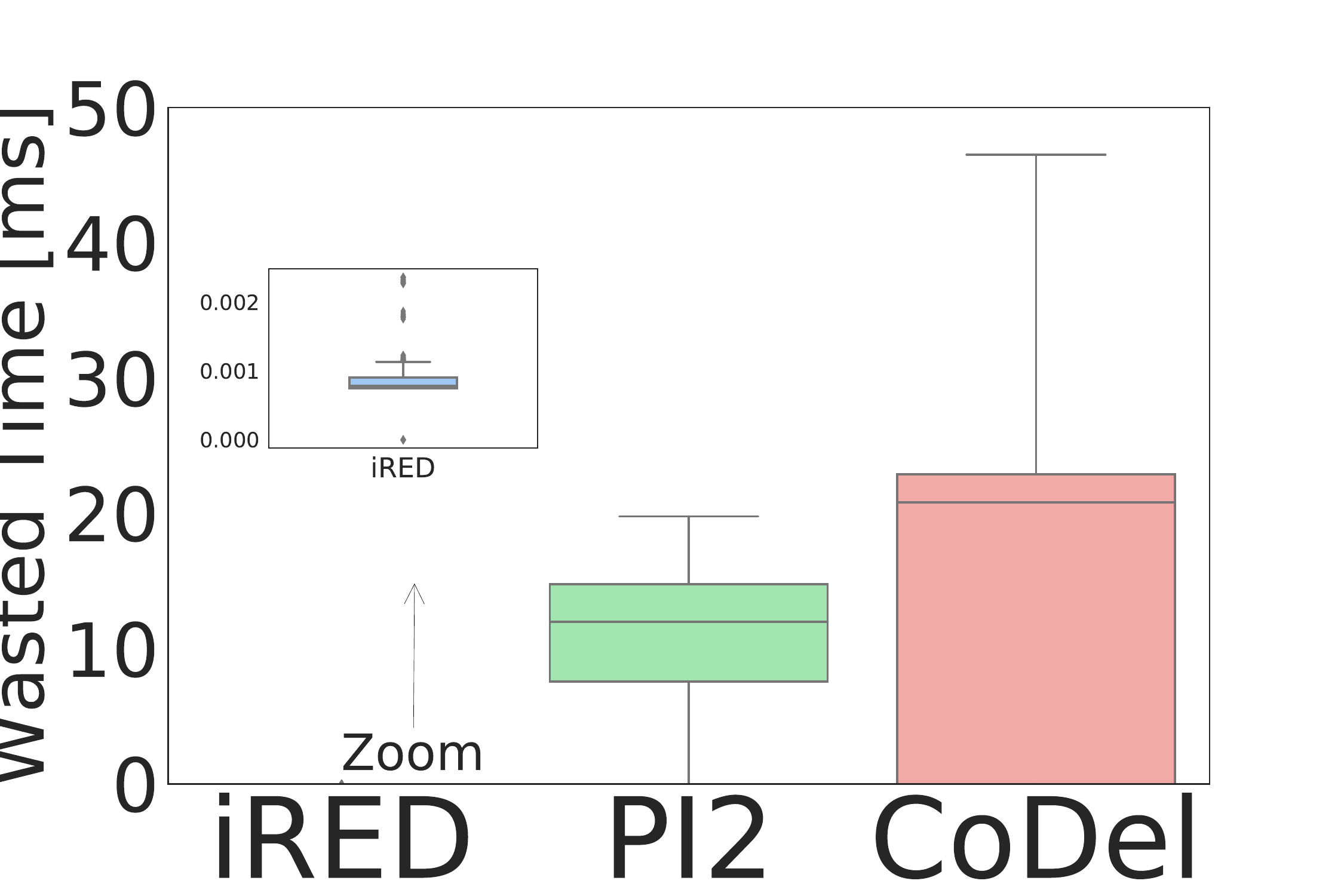}
    }
    \subfigure[VI]{
        \includegraphics[width=.3\columnwidth]{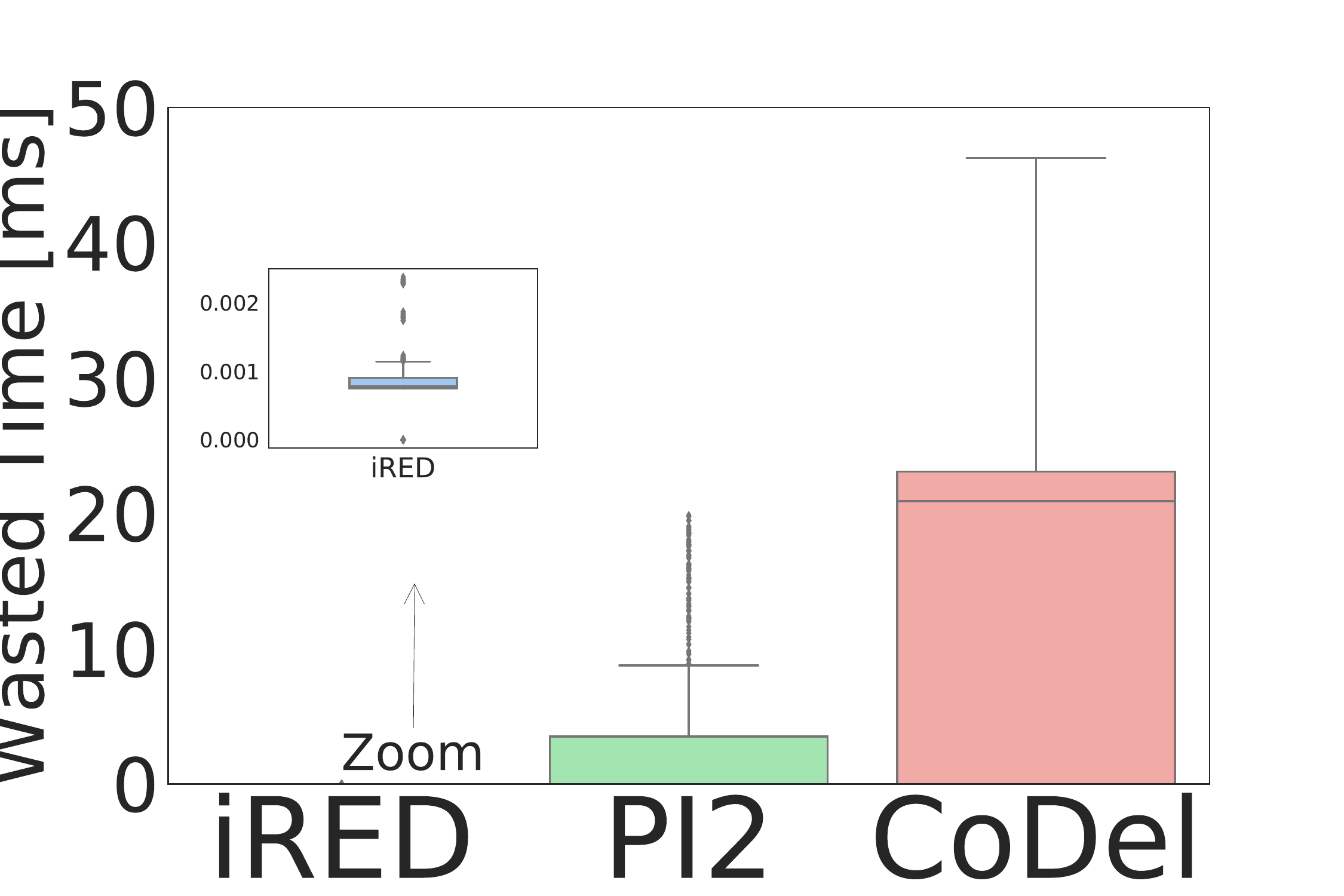}
    } 
    \subfigure[VII]{
        \includegraphics[width=.3\columnwidth]{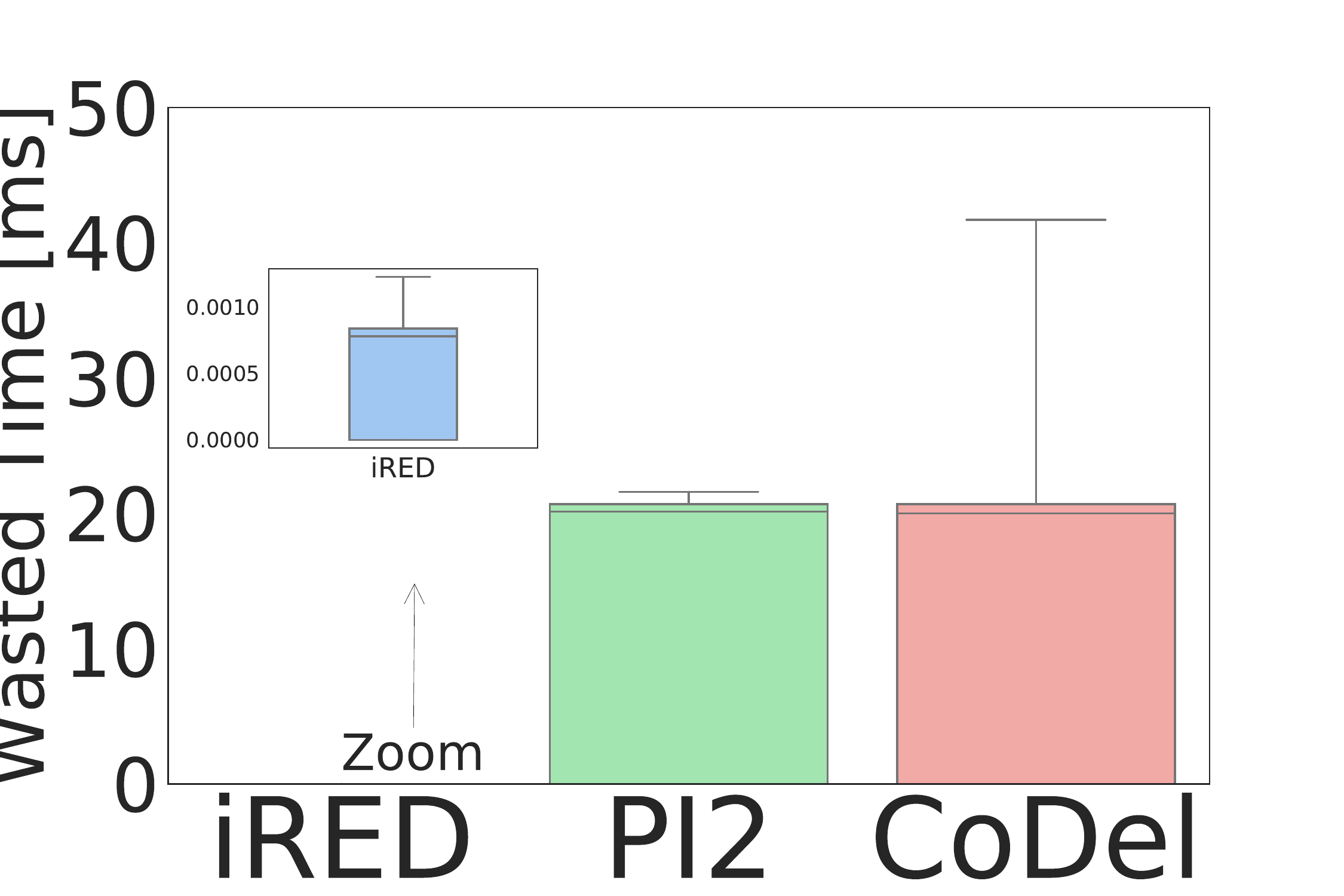}
    }
    \subfigure[VIII]{
        \includegraphics[width=.3\columnwidth]{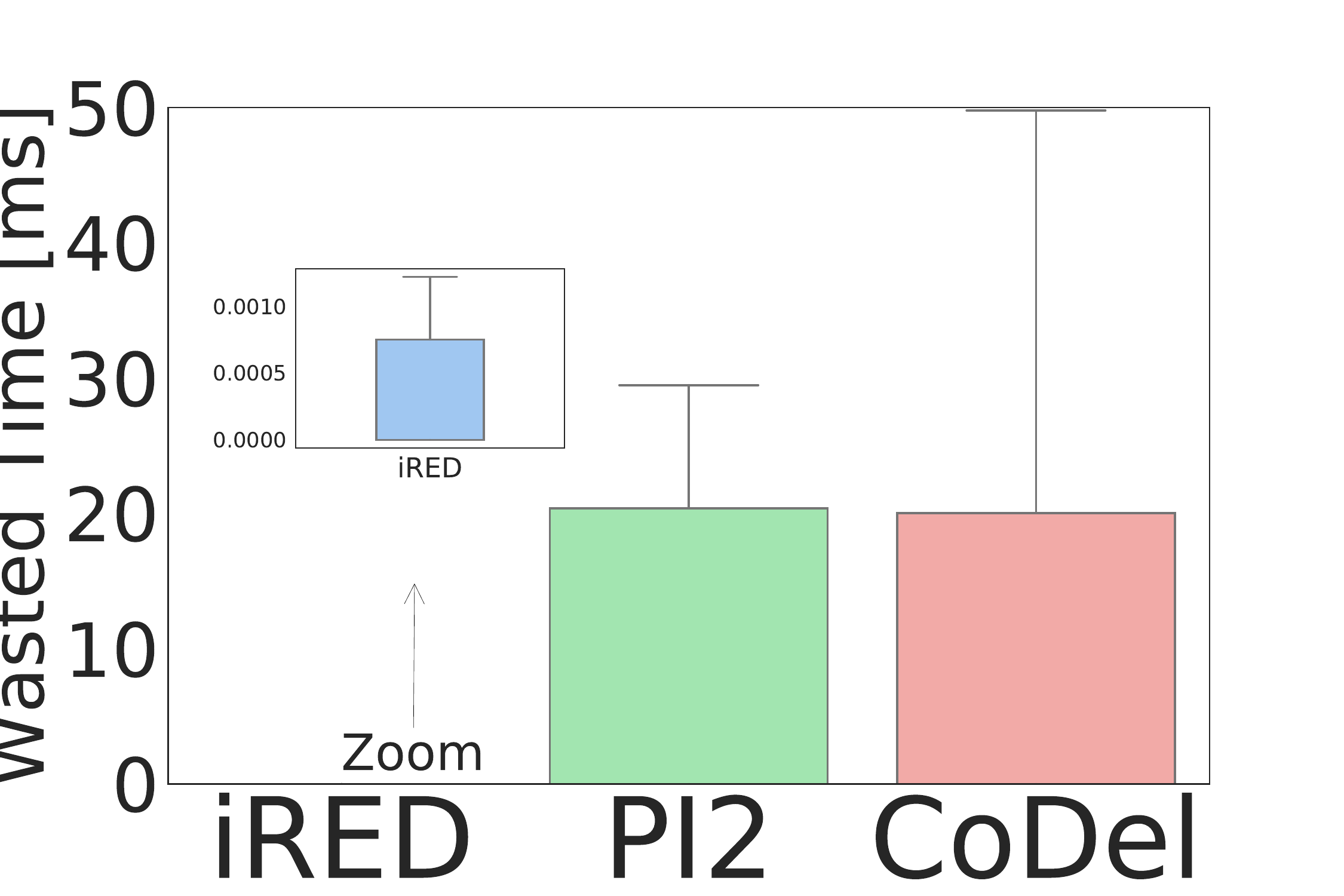}
    }
     \subfigure[IX]{
        \includegraphics[width=.3\columnwidth]{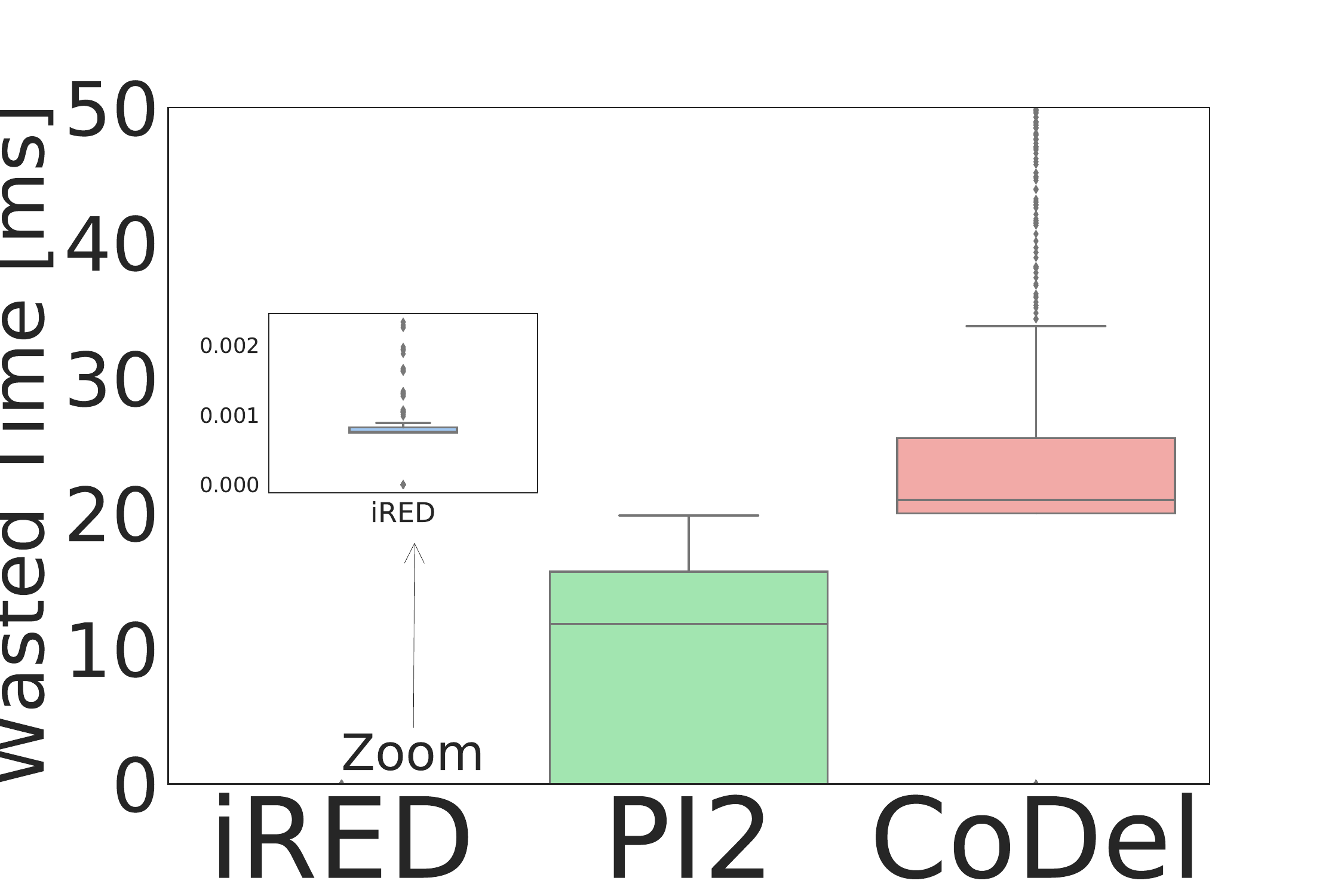}
    }
    \subfigure[X]{
        \includegraphics[width=.3\columnwidth]{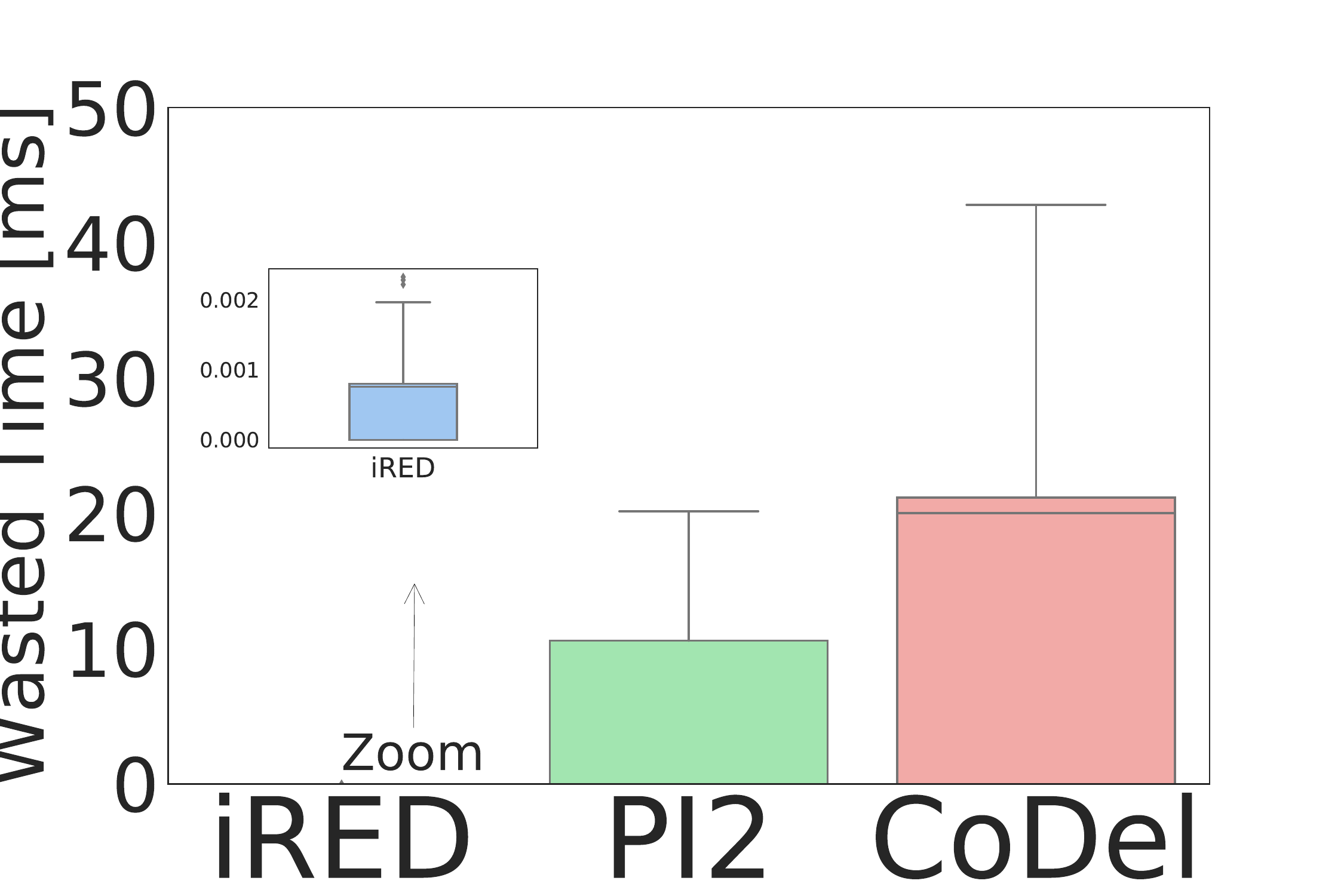}
    }
    \subfigure[XI]{
        \includegraphics[width=.3\columnwidth]{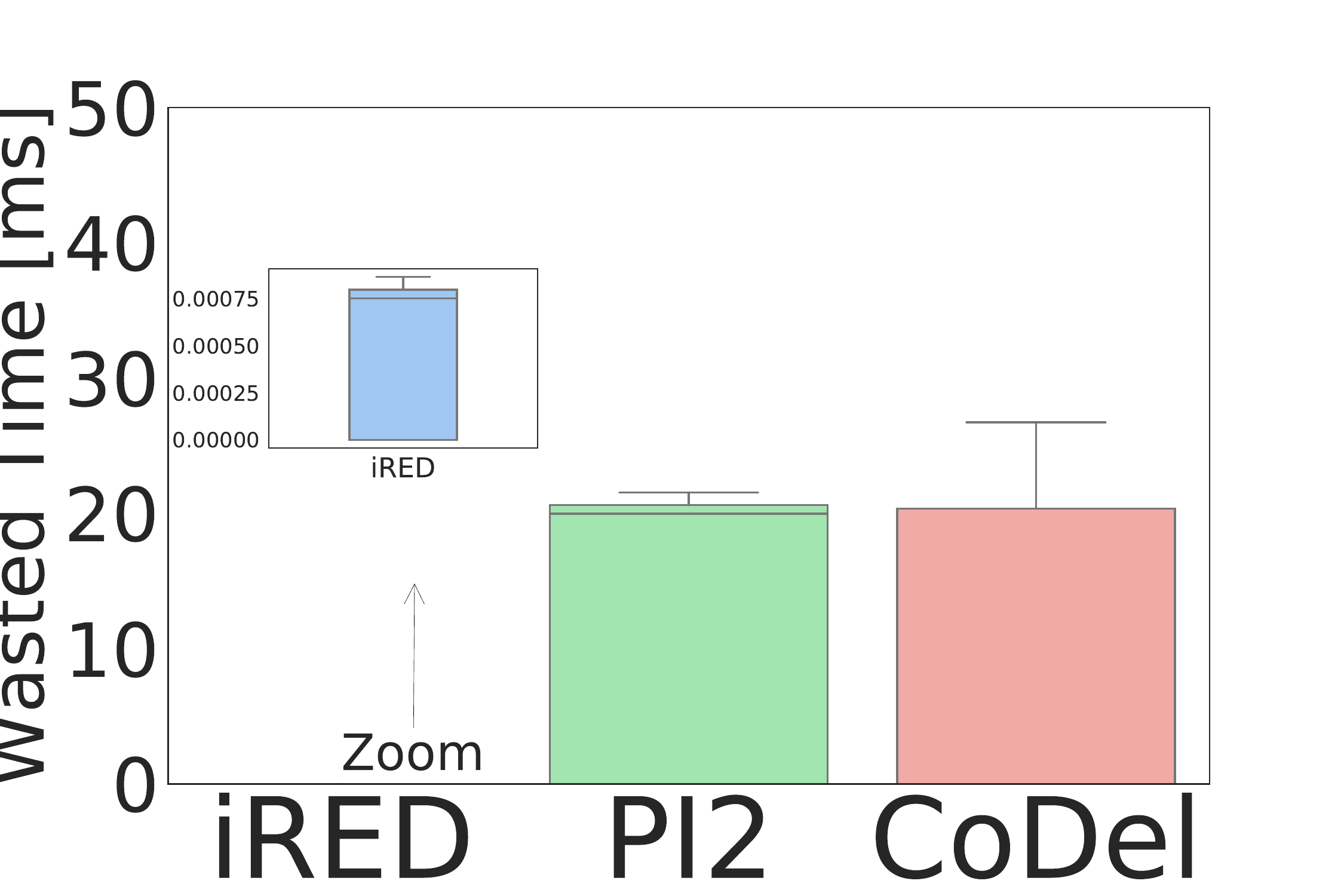}
    }
    \subfigure[XII]{
        \includegraphics[width=.3\columnwidth]{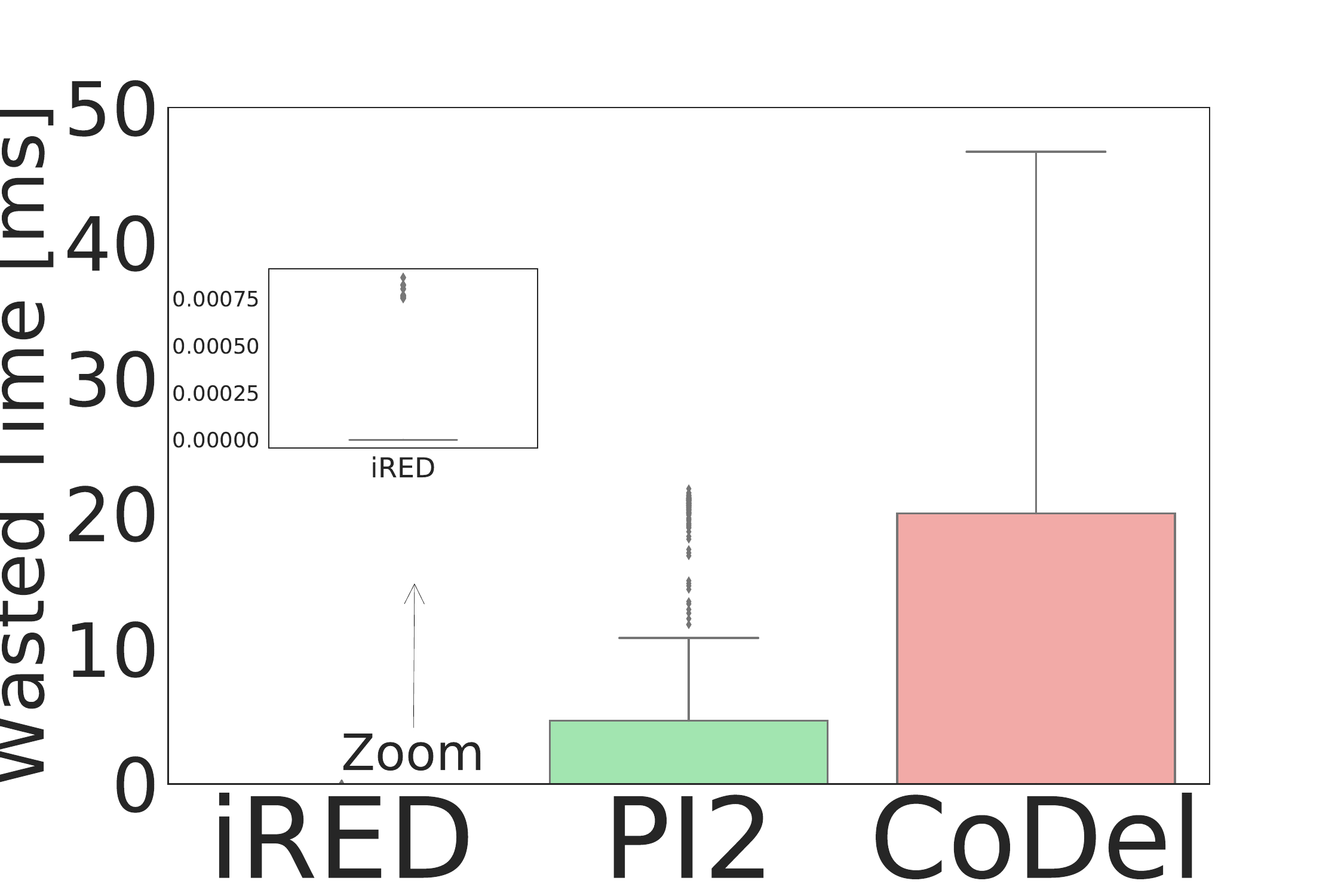}
    }
    \caption{Wasted Resources (Time).}
    \label{fig:WastedResourcesTime}
\end{figure*}

Fig. \ref{fig:WastedResourcesTime} shows the boxplot of the wasted time for iRED, P4-Codel and PI2. For reasons already explained, the iRED+G is not present in this measurement. In many of the observed cases for P4-CoDel and PI2, the median of the wasted time for the discarded packets is very close to the TARGET DELAY, that is, the packets waited in the queue for about 20ms before being discarded.  

On the other hand, for iRED, the wasted time was very low, even requiring a zoom (blue boxplot) in the graph for better visualization of the measurements. In this case, only 48 bytes are transferred when the AQM logic decides to drop, that is, consumes very low time through the 400Gbps internal recirculation port.

Since iRED uses the high-speed recirculation port, the recirculation time is very small compared to the queuing delay of Egress-based AQMs. For example, the recirculation time was approximately 0.001ms per packet in all configurations evaluated, while dropped packets wasted 20ms on Egress-based AQMs. This explains why we need to zoom in on Fig. \ref{fig:WastedResourcesTime}.

\subsubsection{Wasted Latency and Power}

The results shown in this section were obtained using the P4 Insight (\textit{p4i}) tool\footnote{https://www.intel.com.br/content/www/br/pt/products/details/network-io/intelligent-fabric-processors/p4-insight.html.} provided by Intel to inspect the P4-codes. First of all, by means of P4 Insigtht, we obtained the cycles and power consumption for each AQM. Table \ref{tab:WastedPowerandLatency} summarizes the \textit{p4i} output for each metric evaluated (for each programmable block).

The number of Cycles or Weight for iRED is more balanced between Ingress and Egress. Although for iRED the Cycles/Weight numbers are balanced between Ingress and Egress, the dropped packets essentially consumed the resources of the Ingress block. Not surprisingly, for PI2 and CoDel, most of the Cycles and Weights are concentrated in the Egress. Noteworthy to say that, although the number of cycles for PI2 is smaller than CoDel, the weight is larger for PI2. The explanation refers to the fact that  PI2 needs additional registers to store the probabilities computed by the control plane, requiring more power consumption for the writing operations. 
For iRED+G, all AQM logic is concentrated in the Ingress.

\begin{table}[!htpb]
\caption{Latency and Power}
\centering
\label{tab:WastedPowerandLatency}
\begin{tabular}{ccccccccc}
\toprule
 \multirow{2}{*}{\textbf{AQMs}}& \multicolumn{2}{c}{\textbf{Cycles}} & \multicolumn{2}{c}{\textbf{Weight}} \\ 
 & \textbf{Ingress} & \textbf{Egress} & \textbf{Ingress} & \textbf{Egress}\\
\midrule
\textbf{iRED} & \cellcolor[gray]{0.7591}108.0 & \cellcolor[gray]{0.5888}192.0 & \cellcolor[gray]{0.7499}112.5 & \cellcolor[gray]{0.6561}158.8 \\
\textbf{PI2} & \cellcolor[gray]{0.8564}60.0 & \cellcolor[gray]{0.6536}160.0 & \cellcolor[gray]{0.9358}20.8 & \cellcolor[gray]{0.5000}235.8 \\
\textbf{CoDel} & \cellcolor[gray]{0.8564}60.0 & \cellcolor[gray]{0.5807}196.0 & \cellcolor[gray]{0.9500}13.8 & \cellcolor[gray]{0.6640}154.9 \\
\textbf{iRED+G} & \cellcolor[gray]{0.5482}212.0 & \cellcolor[gray]{0.8077}84.0 & \cellcolor[gray]{0.5564}208.0 & \cellcolor[gray]{0.9139}31.6 \\
\bottomrule
\end{tabular}
\end{table}

Then, by having the numbers shown in Tab. \ref{tab:WastedPowerandLatency}, we were able to calculate the wasted cycles and weight. 

\subsubsection{Wasted Clock Cycles}

%The Wasted Cycles represent the sum of clock cycles consumed (Ingress and Egress) by each dropped packet. 
Each block has a fixed number of clock cycles (Latency), which are necessary to forward each packet through the pipeline. For PI2, the wasted cycles are computed by $60 + 160 = 220$ cycles per dropped packet. For P4-CoDel, the wasted cycles are computed by $60 + 196 = 256$ cycles per dropped packet. In iRED and iRED+G cases, only Ingress cycles are used, resulting in 108 and 212 per dropped packet respectively. 

\begin{table}[!htpb]
\centering
\caption{Wasted Clock Cycles}
\label{tab:cycles}
\begin{tabular}{lcccccccc}
\toprule
\textbf{Conf} & \textbf{iRED} & \textbf{PI2} & \textbf{CoDel} & \textbf{iRED+G} \\
\midrule
\textbf{I} & \cellcolor[gray]{0.8238}3844476.0 & \cellcolor[gray]{0.5000}12851740.0 & \cellcolor[gray]{0.6320}9180416.0 & \cellcolor[gray]{0.6757}7962932.0 \\
\textbf{II} & \cellcolor[gray]{0.9180}1221588.0 & \cellcolor[gray]{0.7822}5001700.0 & \cellcolor[gray]{0.7968}4594432.0 & \cellcolor[gray]{0.8737}2453900.0 \\
\textbf{III} & \cellcolor[gray]{0.9251}1025460.0 & \cellcolor[gray]{0.9023}1660120.0 & \cellcolor[gray]{0.6251}9370112.0 & \cellcolor[gray]{0.6135}9693700.0 \\
\textbf{IV} & \cellcolor[gray]{0.9460}443124.0 & \cellcolor[gray]{0.9477}396440.0 & \cellcolor[gray]{0.8783}2326016.0 & \cellcolor[gray]{0.8609}2812392.0 \\
\textbf{V} & \cellcolor[gray]{0.8539}3005208.0 & \cellcolor[gray]{0.6610}8373200.0 & \cellcolor[gray]{0.7661}5448192.0 & \cellcolor[gray]{0.7408}6151392.0 \\
\textbf{VI} & \cellcolor[gray]{0.9323}823500.0 & \cellcolor[gray]{0.8098}4234120.0 & \cellcolor[gray]{0.7718}5290240.0 & \cellcolor[gray]{0.9004}1712324.0 \\
\textbf{VII} & \cellcolor[gray]{0.9381}663228.0 & \cellcolor[gray]{0.9261}998360.0 & \cellcolor[gray]{0.6570}8482816.0 & \cellcolor[gray]{0.7458}6013804.0 \\
\textbf{VIII} & \cellcolor[gray]{0.9336}787968.0 & \cellcolor[gray]{0.9431}523160.0 & \cellcolor[gray]{0.8211}3917056.0 & \cellcolor[gray]{0.8430}3309744.0 \\
\textbf{IX} & \cellcolor[gray]{0.8908}1977912.0 & \cellcolor[gray]{0.7511}5865860.0 & \cellcolor[gray]{0.6579}8459264.0 & \cellcolor[gray]{0.7955}4630292.0 \\
\textbf{X} & \cellcolor[gray]{0.9408}589140.0 & \cellcolor[gray]{0.8682}2607440.0 & \cellcolor[gray]{0.8660}2670080.0 & \cellcolor[gray]{0.9164}1266276.0 \\
\textbf{XI} & \cellcolor[gray]{0.9438}505872.0 & \cellcolor[gray]{0.9411}580580.0 & \cellcolor[gray]{0.7413}6137600.0 & \cellcolor[gray]{0.8273}3745828.0 \\
\textbf{XII} & \cellcolor[gray]{0.9150}1304640.0 & \cellcolor[gray]{0.9500}332200.0 & \cellcolor[gray]{0.6871}7646720.0 & \cellcolor[gray]{0.8115}4185728.0 \\
\bottomrule
\end{tabular}
\end{table}

In Table \ref{tab:cycles}, the cycles consumed by iRED for the dropped packets are colored on a lighter scale in most parts of the configurations. If we look at the values, iRED achieves savings in the order of up to 12x fewer clock cycles. Moreover, the results of the iRED+G show that despite running in the Ingress, it wastes a large number of clock cycles for each dropped packet since all AQM logic operations are combined within the same programmable block.

\subsubsection{Wasted Weight (Power Consumption)}

%Weight is a unit-less value that represents the Power consumption in each block of Ingress and Egress (see Table \ref{tab:WastedPowerandLatency}). 
The Wasted Weight is a sum of weights (Power consumption) in Ingress and Egress for each dropped packet. For PI2, the wasted weight is computed by $20.8 + 235.8 = 256,8$ per dropped packet. For P4-CoDel, the wasted weight is computed by $13.8 + 154.9 = 168,7$ per dropped packet. In iRED and iRED+G cases, only Ingress weights are used, resulting in 112.5 and 208 per dropped packet respectively.

\begin{table}[!htpb]
\centering
\caption{Wasted Weight (Power Consumption)}
\label{tab:WastedPower}
\begin{tabular}{lcccccccc}
\toprule
\textbf{Conf} & \textbf{iRED} & \textbf{PI2} & \textbf{CoDel} & \textbf{iRED+G} \\
\midrule
\textbf{I} & \cellcolor[gray]{0.8383}4004662.5 & \cellcolor[gray]{0.5000}14989802.2 & \cellcolor[gray]{0.7753}6049750.7 & \cellcolor[gray]{0.7210}7812688.0 \\
\textbf{II} & \cellcolor[gray]{0.9224}1272487.5 & \cellcolor[gray]{0.7869}5674656.0 & \cellcolor[gray]{0.8684}3027658.9 & \cellcolor[gray]{0.8875}2407600.0 \\
\textbf{III} & \cellcolor[gray]{0.9287}1068187.5 & \cellcolor[gray]{0.9036}1883481.6 & \cellcolor[gray]{0.7715}6174757.4 & \cellcolor[gray]{0.6687}9510800.0 \\
\textbf{IV} & \cellcolor[gray]{0.9474}461587.5 & \cellcolor[gray]{0.9478}449779.2 & \cellcolor[gray]{0.9144}1532808.2 & \cellcolor[gray]{0.8766}2759328.0 \\
\textbf{V} & \cellcolor[gray]{0.8652}3130425.0 & \cellcolor[gray]{0.6691}9499776.0 & \cellcolor[gray]{0.8510}3590273.4 & \cellcolor[gray]{0.7758}6035328.0 \\
\textbf{VI} & \cellcolor[gray]{0.9352}857812.5 & \cellcolor[gray]{0.8137}4803801.6 & \cellcolor[gray]{0.8543}3486185.5 & \cellcolor[gray]{0.9099}1680016.0 \\
\textbf{VII} & \cellcolor[gray]{0.9403}690862.5 & \cellcolor[gray]{0.9267}1132684.8 & \cellcolor[gray]{0.7895}5590043.2 & \cellcolor[gray]{0.7799}5900336.0 \\
\textbf{VIII} & \cellcolor[gray]{0.9363}820800.0 & \cellcolor[gray]{0.9433}593548.8 & \cellcolor[gray]{0.8821}2581278.7 & \cellcolor[gray]{0.8616}3247296.0 \\
\textbf{IX} & \cellcolor[gray]{0.8982}2060325.0 & \cellcolor[gray]{0.7567}6655084.8 & \cellcolor[gray]{0.7899}5574522.8 & \cellcolor[gray]{0.8217}4542928.0 \\
\textbf{X} & \cellcolor[gray]{0.9427}613687.5 & \cellcolor[gray]{0.8705}2958259.2 & \cellcolor[gray]{0.9074}1759541.0 & \cellcolor[gray]{0.9233}1242384.0 \\
\textbf{XI} & \cellcolor[gray]{0.9454}526950.0 & \cellcolor[gray]{0.9413}658694.4 & \cellcolor[gray]{0.8371}4044582.5 & \cellcolor[gray]{0.8484}3675152.0 \\
\textbf{XII} & \cellcolor[gray]{0.9198}1359000.0 & \cellcolor[gray]{0.9500}376896.0 & \cellcolor[gray]{0.8064}5039069.0 & \cellcolor[gray]{0.8351}4106752.0 \\
\bottomrule
\end{tabular}
\end{table}

Looking at Table \ref{tab:WastedPower}, the Egress-based AQMs have more power consumption in comparison to iRED, because all drop logic is not disaggregated. This is repeated with the iRED+G version, which concentrates all operations in the Ingress block. On the other hand, as iRED splits AQM's operations, only the Ingress block's power resources are consumed by dropped packets. Then, iRED reduces power consumption by up to 8x.

\subsubsection{Consolidation of the results}

\begin{figure}[ht]
    \centering
    \includegraphics[width=.8\columnwidth]{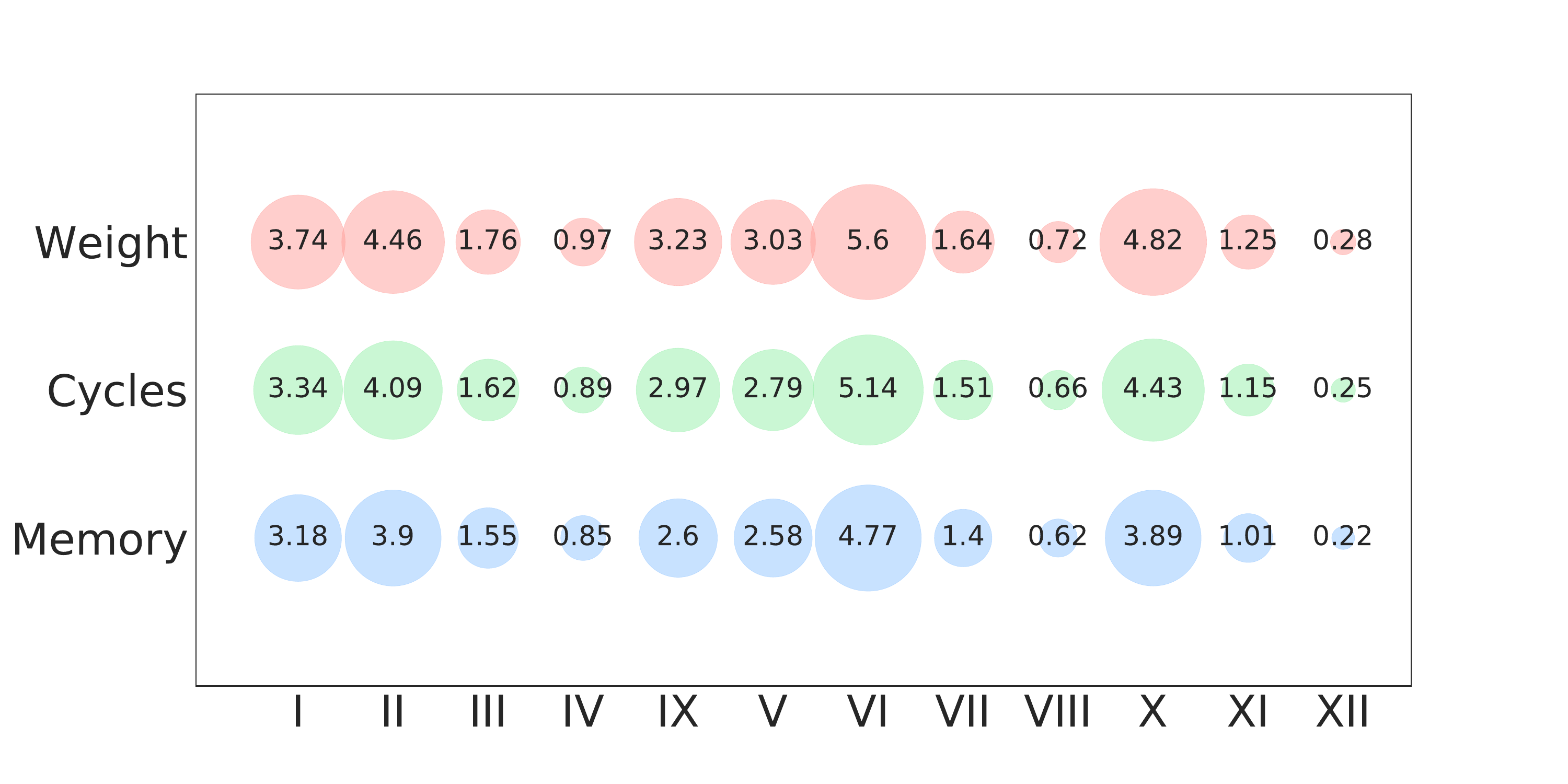}
    \caption{iRED resource savings compared to PI2.}
    \label{fig:PI2-consolidation}
\end{figure}

\begin{figure}[ht]
    \centering
    \includegraphics[width=.8\columnwidth]{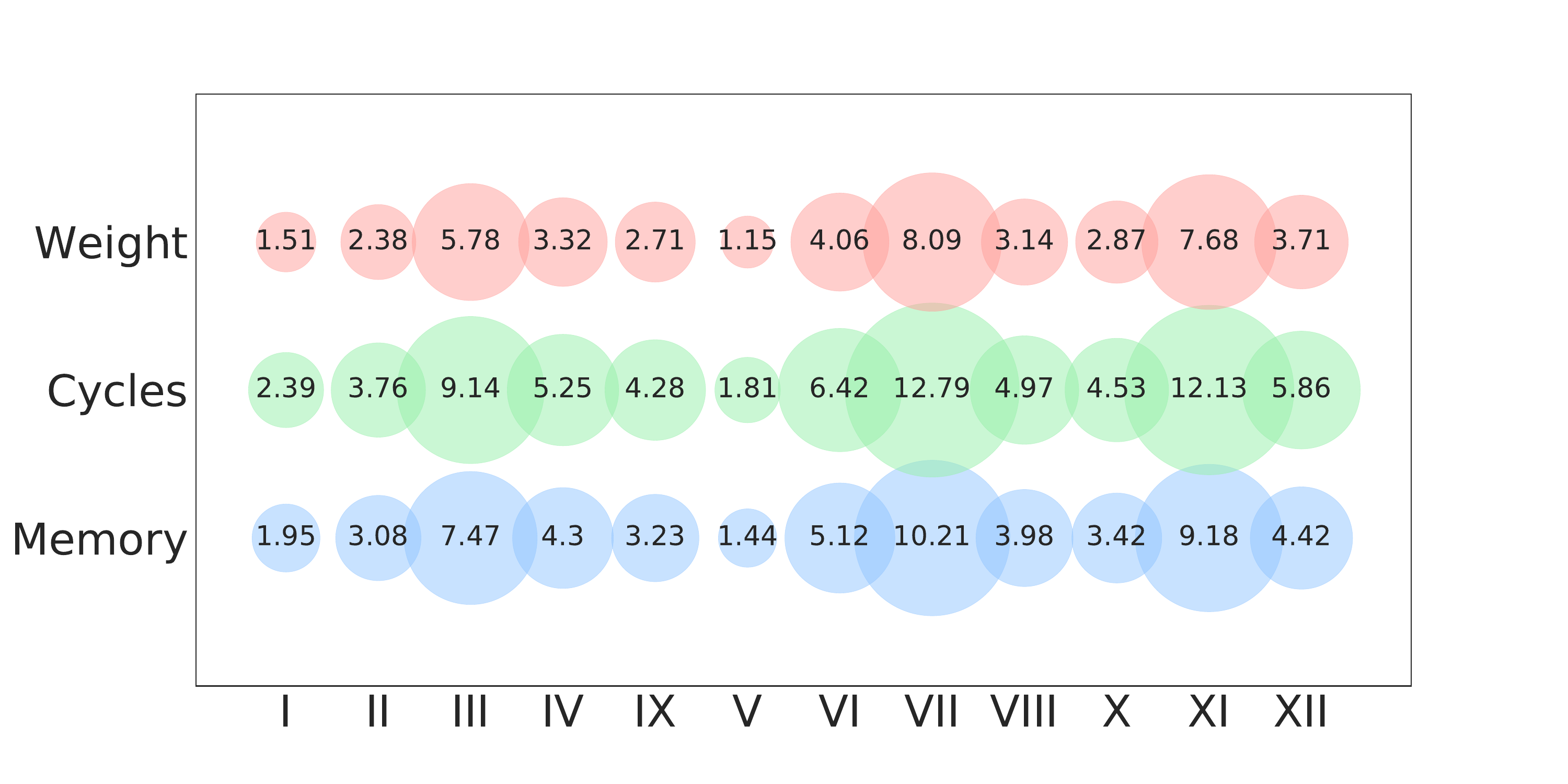}
    \caption{iRED resource savings compared to P4-CoDel.}
    \label{fig:CoDel-consolidation}
\end{figure}

Figs. \ref{fig:PI2-consolidation} and \ref{fig:CoDel-consolidation} show the consolidated overview of the resources saved by iRED for all configurations evaluated. Regarding PI2 (Fig. \ref{fig:PI2-consolidation}), iRED saves up to 5.6x power consumption, 5.47x clock cycles and 4.77x memory. However, in three configurations (IV, VIII and XII) in which the RTT is 50ms, PI2 wastes fewer resources. We observed that the target delay was rarely reached in these configurations, resulting in few actions of the PI2. %The delay generated by the control plane, added to the delay emulated in the network, caused this behavior.}

Regarding P4-CoDel (Fig. \ref{fig:CoDel-consolidation}), the AQM algorithm drops all packets that reached the target delay. Even for the scalable traffic (TCP Prague), all packets are dropped (instead of being marked), since P4-CoDel does not support the L4S. This explains the large number of wasted resources. iRED saves up to 8.9x weight, 12.79x clock cycles and 10.21x memory. 

%\subsection{Lessons learned and takeaways}

%\textcolor{red}{To meet their goals, the AQMs need drop packets. However, the discussion here is not on the fact of dropping more or fewer packets, but on how many resources can be saved to accomplish this task. To understand this fact, we need to know how and mainly where this discard is carried out. PDP hardware allows us some flexibility compared to fixed-function ASICs. However, it still imposes several challenges and constraints intrinsic to each architecture.}

\subsection{Fair sharing in L4S scope}

In this particular experiment, our primary objective is to evaluate the extent of support and adherence to the L4S framework. As only iRED and PI2 meet this requirement, P4-CoDel was not evaluated in this experiment. Additionally, we aim to evaluate the harmonious coexistence between non-L4S flows, conventionally referred to as classic (TCP Cubic), and L4S-compliant flows, denoted as Scalable (TCP Prague). We used the same setup as the previous experiment (See Fig. \ref{fig:setup}), selecting configurations with an MTU of 1500 bytes (I, II, III, and IV).

In alignment with the methodology outlined by \cite{DualPI2}, our experimental configuration involved the imposition of traffic intensity loads comprising four discrete phases, each spanning a 120-second duration. Within each of these phases, we introduced new flows with specific flow pairs (1-1, 2-2, 10-10, 25-25) into the system. This sequential introduction of flows allowed us to initiate the load with lower intensity and progress toward a high-load scenario.

\begin{figure}[ht]
    \centering
    \subfigure[iRED 120Mbps, 10ms]{
        \includegraphics[width=.4\columnwidth]{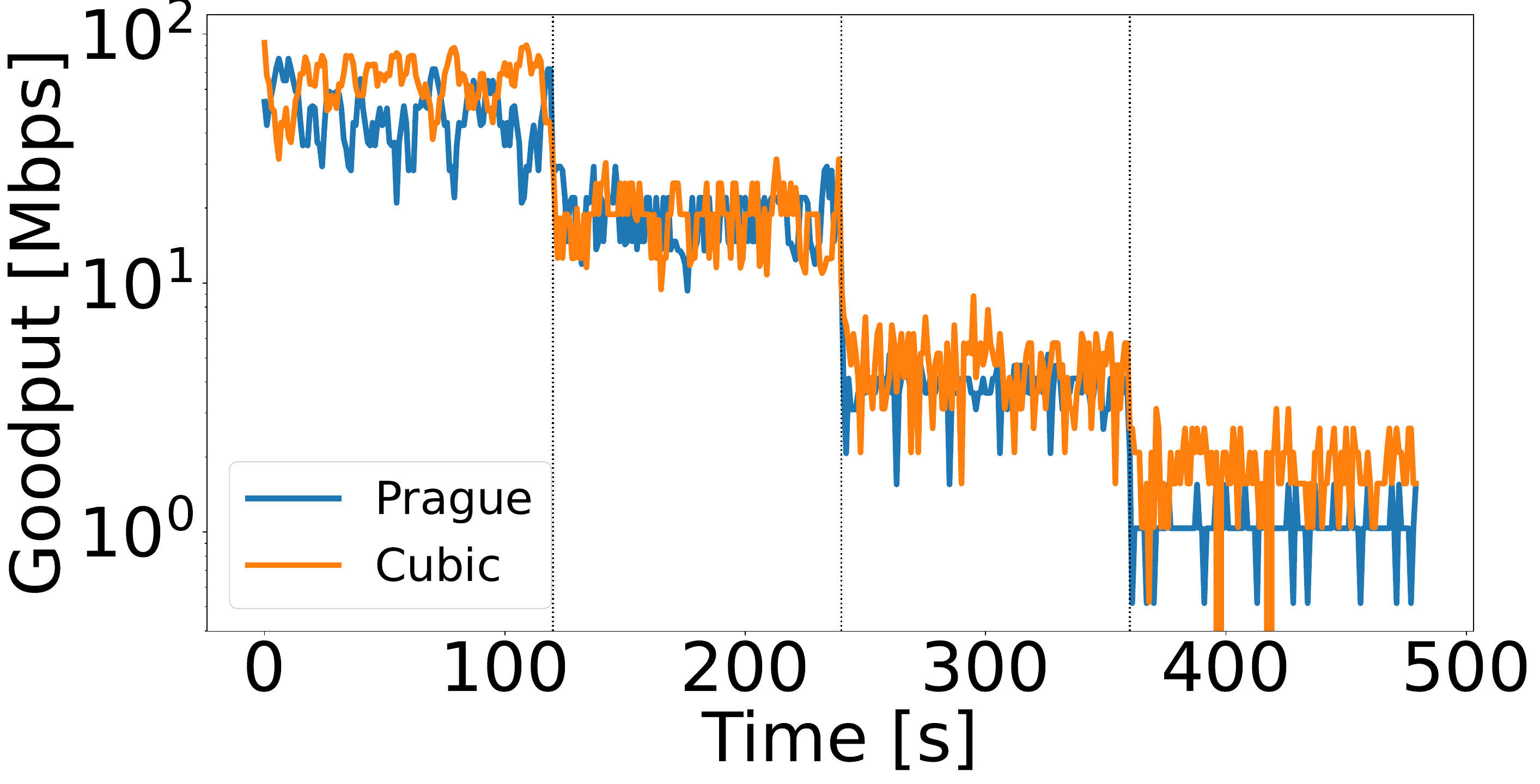}
        \label{fig:ired12010}
    }
    \subfigure[PI2 120Mbps, 10ms]{
       \includegraphics[width=.4\columnwidth]{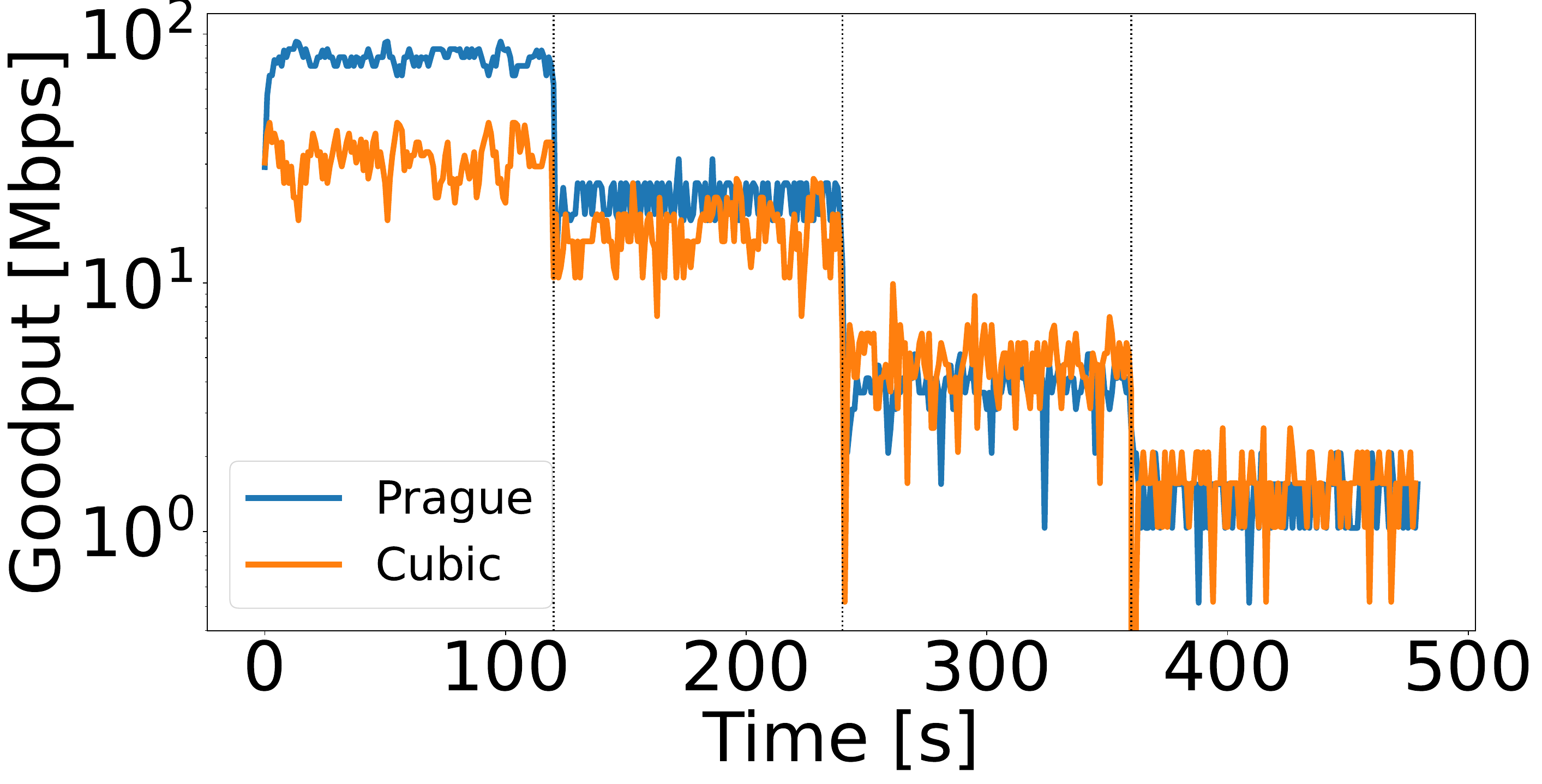}
       \label{fig:pi212010}
    }
       \subfigure[iRED 1Gbps, 10ms]{
        \includegraphics[width=.4\columnwidth]{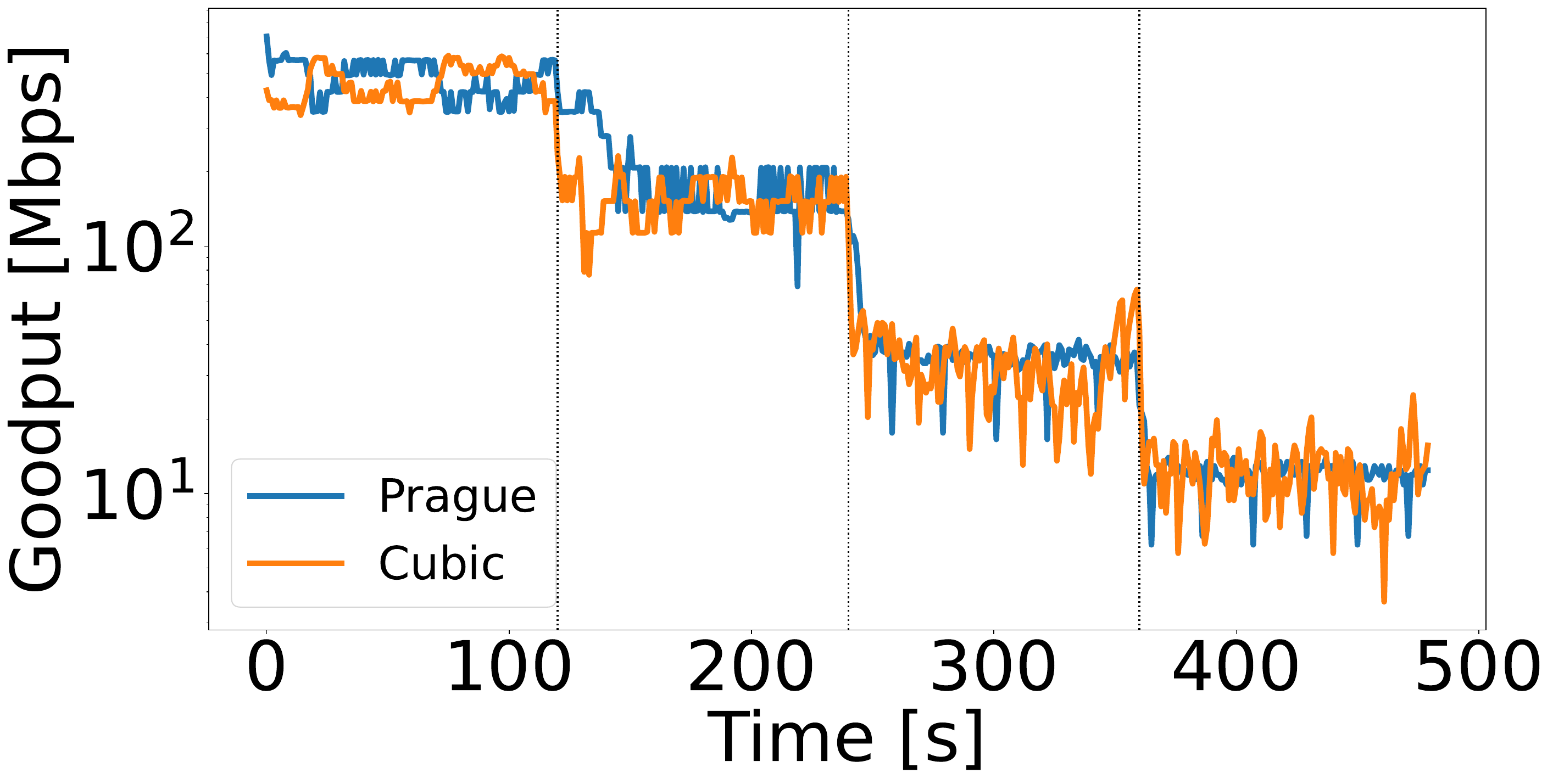}
        \label{fig:ired100010}
    }
    \subfigure[PI2 1Gbps, 10ms]{
         \includegraphics[width=.4\columnwidth]{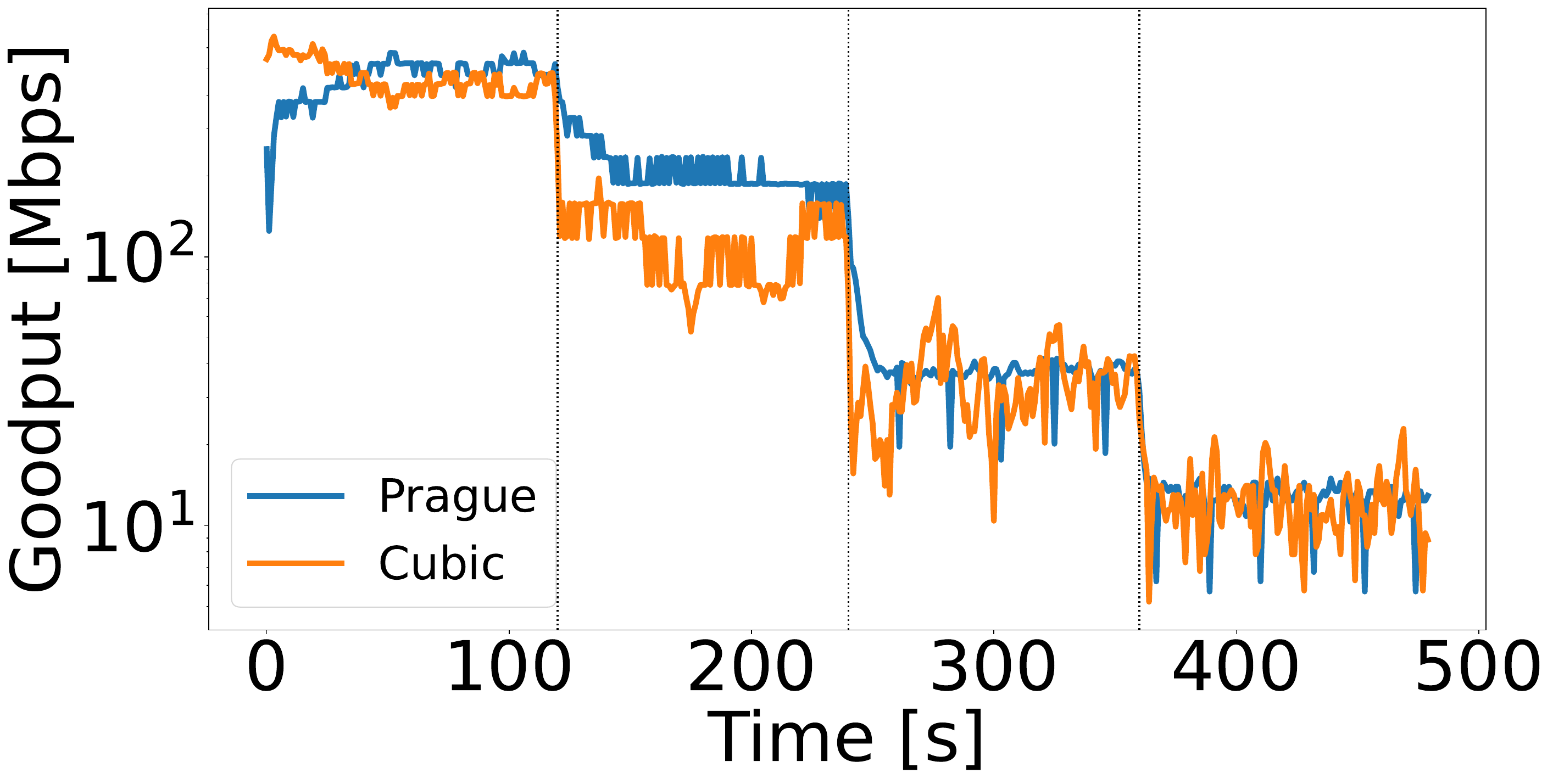}
         \label{fig:pi2100010}
    }
    \caption{Coexistence evaluation of Cubic and Prague flows (RTT base 10ms).}
    \label{fig:RTT10ms}
\end{figure}

In the context of a 10ms baseline RTT and a bandwidth set at 120 Mbps, Figs. \ref{fig:ired12010} and \ref{fig:pi212010}, becomes evident that a more equitable coexistence between flows is achieved with the implementation of the iRED. Conversely, flows employing the PI2 exhibit a relative disadvantage, with improved fairness only becoming apparent in the latter half of the experiment, specifically during phases 3 and 4.

When we examine the evaluation outcomes for a 1 Gbps bandwidth and a base RTT of 10 ms, it remains evident that the equitable distribution of shared bandwidth among flows persists across all phases of the experiment when utilizing iRED, as can be seen in Fig. \ref{fig:ired100010}. In the case of PI2, Fig. \ref{fig:pi2100010} despite the initial appearance of fairness in the coexistence of flows during the initial phase of the experiment, this equilibrium does not endure into phase 2.

In Fig. \ref{fig:RTT10ms}, the overarching conclusion drawn from our analysis suggests that in the case of PI2, the intensity (i.e., the probability of marking the ECN bit) required to mark packets from the Prague flow is insufficient during the initial phases of the experiment. This deficiency in marking intensity becomes apparent because the Prague flow, due to its bandwidth consumption characteristics, tends to dominate and not facilitate a fair coexistence with the Cubic flow.

\begin{figure}[ht]
    \centering
    \subfigure[iRED 120Mbps, 50ms]{
        \includegraphics[width=.4\columnwidth]{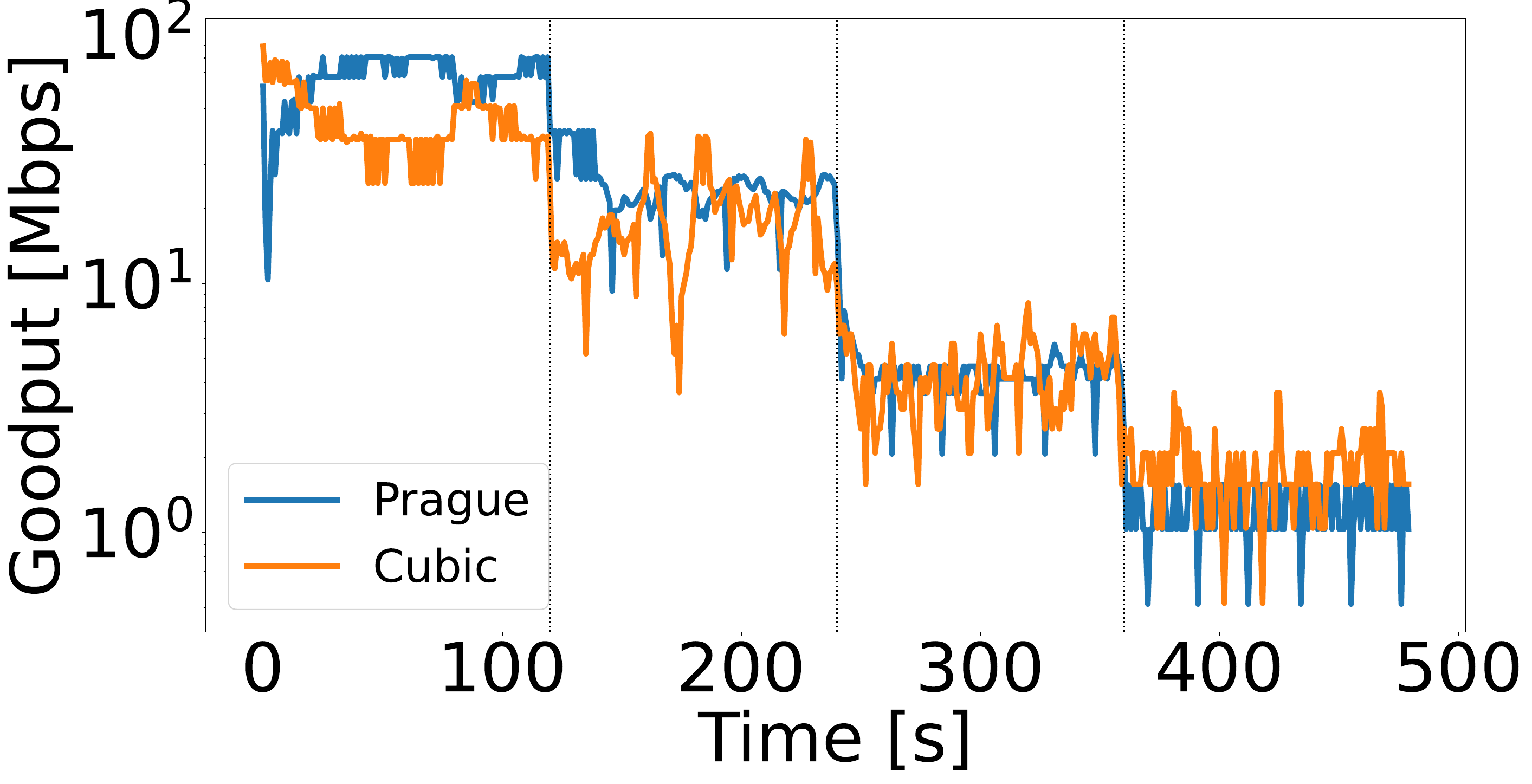}
        \label{fig:ired12050}
    }
    \subfigure[PI2 120Mbps, 50ms]{
        \includegraphics[width=.4\columnwidth]{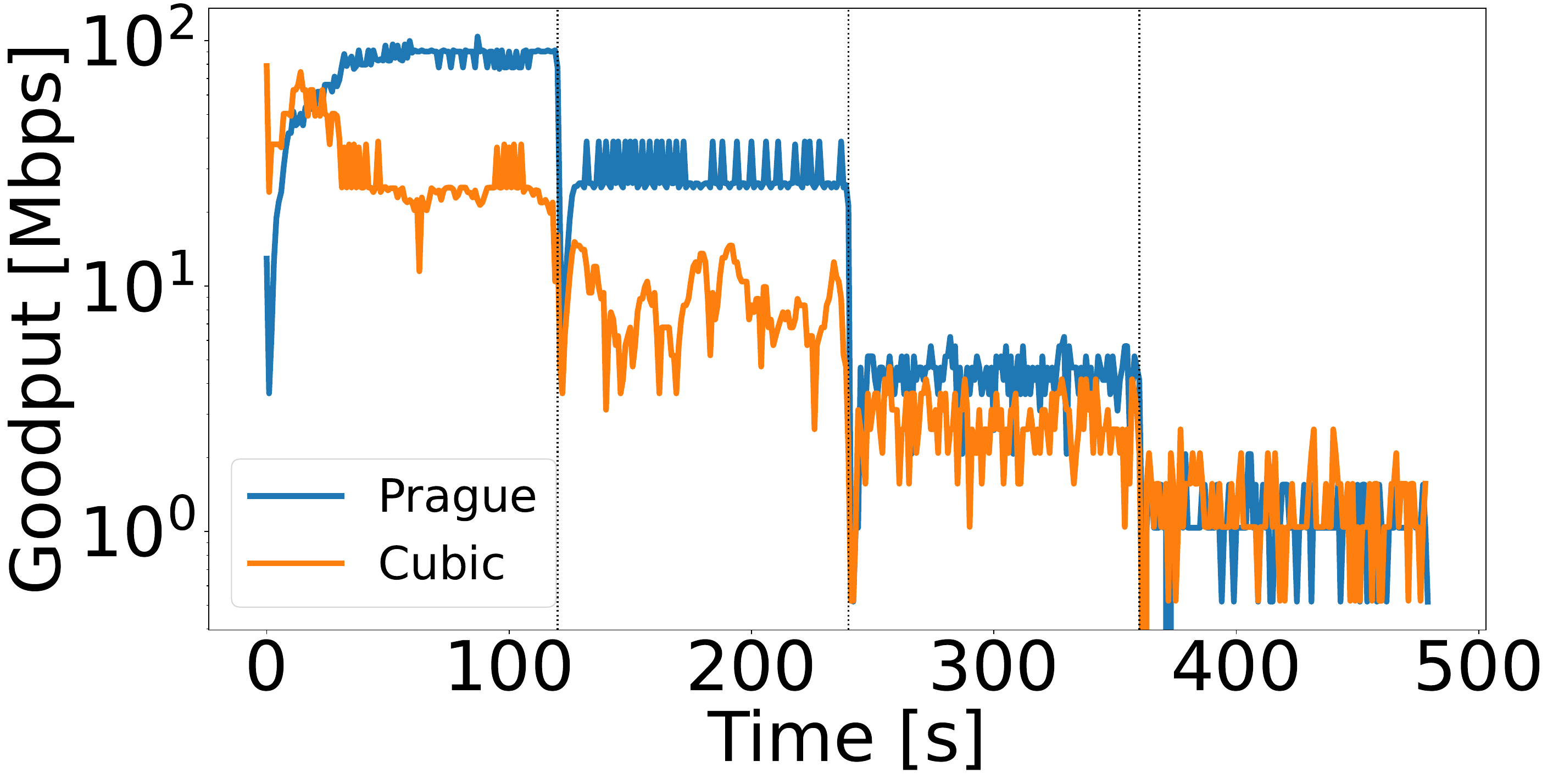}
        \label{fig:pi212050}
    }
       \subfigure[iRED 1Gbps, 50ms]{
        \includegraphics[width=.4\columnwidth]{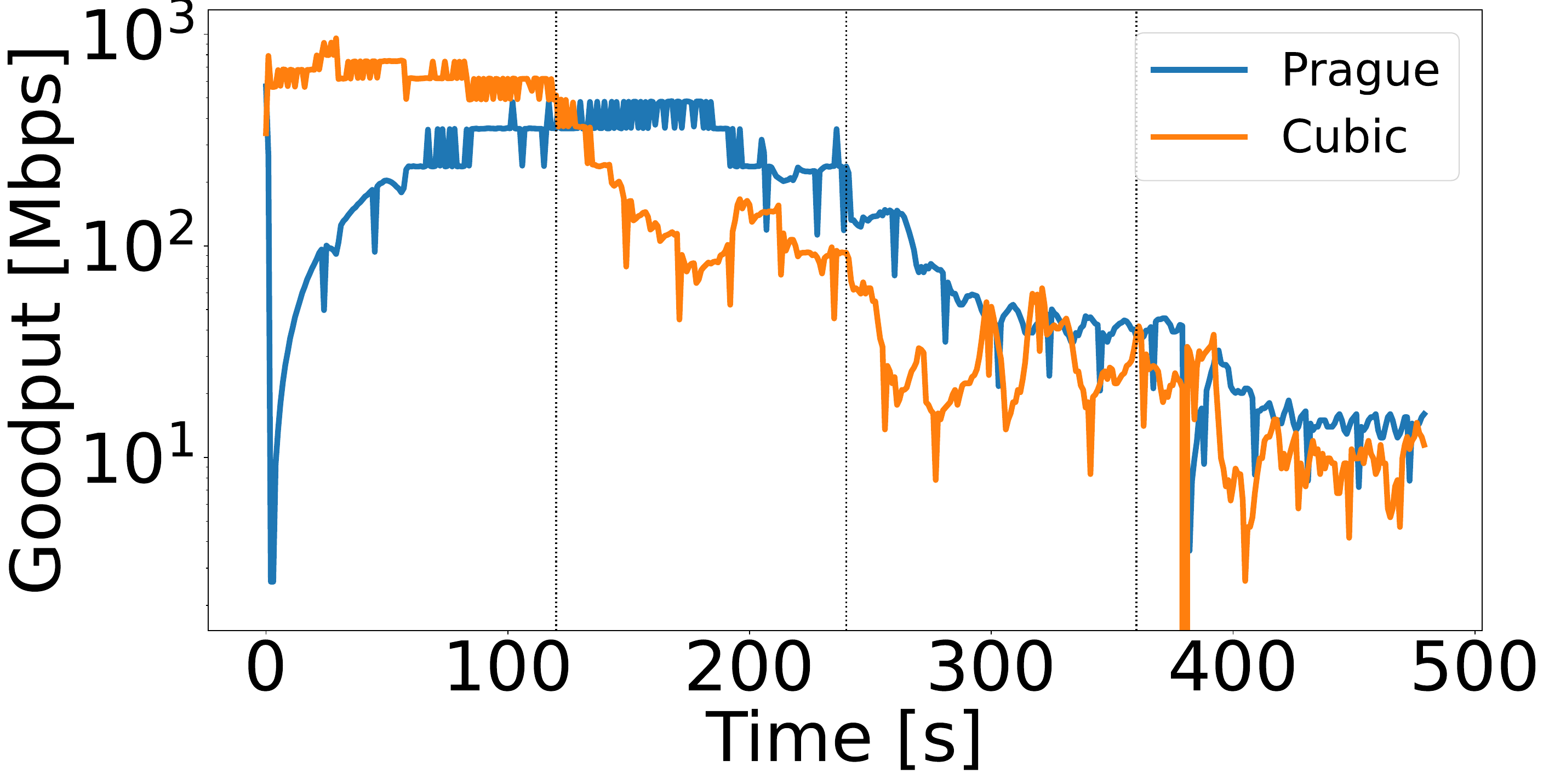}
        \label{fig:ired100050}
    }
    \subfigure[PI2 1Gbps, 50ms]{
        \includegraphics[width=.4\columnwidth]{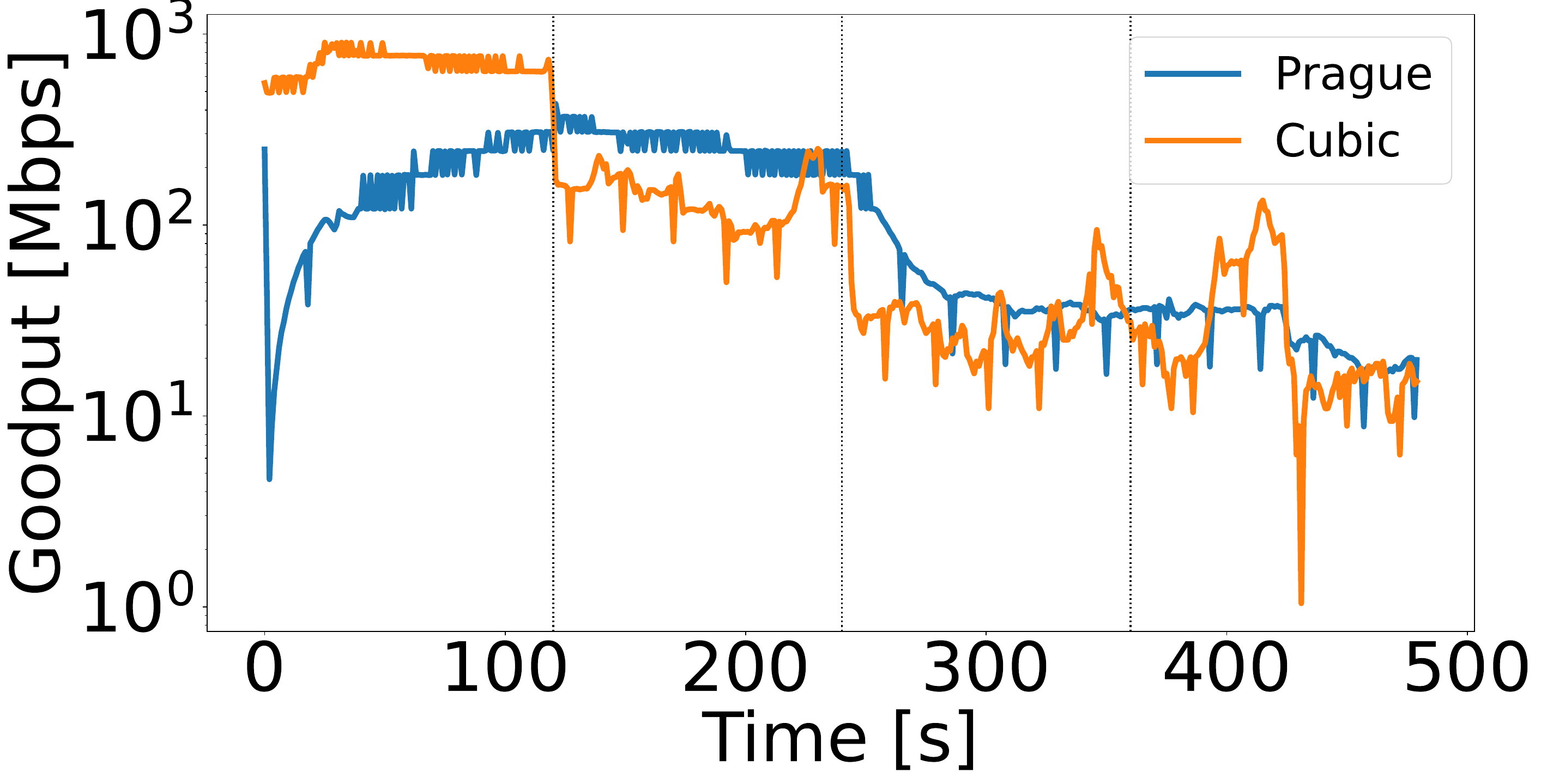}
        \label{fig:pi2100050}
    }
    \caption{Coexistence evaluation of Cubic and Prague flows (RTT base 50ms)}
    \label{fig:RTT50ms}
\end{figure}

In Figure \ref{fig:RTT50ms}, we assess scenarios in which the baseline RTT is configured to 50 ms, a value commonly encountered in long-distance networks. With a bandwidth of 120 Mbps and an RTT of 50 ms, the observed outcomes closely parallel those obtained with an RTT of 10 ms. Specifically, the iRED continues to exhibit superior fairness in the coexistence of Cubic and Prague flows, as seen in Fig. \ref{fig:ired12050}, while the PI2 attains fairness only in the later stages of the experiment, as can be seen in Fig. \ref{fig:pi212050}.

However, in the case of 1 Gbps and an RTT of 50 ms, both approaches exhibited a parallel pattern of behavior, as can be seen in Figs. \ref{fig:ired100050} and \ref{fig:pi2100050}. There was a notable reduction in the performance of the Prague flow during the initial phase of the experiment, followed by a more equitable coexistence between flows in the subsequent three phases. In this particular scenario, our conjecture is that the delayed feedback (ACK) to the Prague TCP flow resulted in a slower initial ramp-up, as Prague TCP is notably more dependent on this metric \cite{Briscoe2018ImplementingT}. This sensitivity likely contributed to the observed behavior where Prague TCP experienced a significant drop in performance during the initial phase of the experiment.

\subsection{DASH scenario}

Finally, there is nothing more important than evaluating novel mechanisms using real scenarios and applications. In this experiment, we elucidate the functioning of delay-based AQM mechanisms, specifically P4-CoDel and PI2, in conjunction with iRED depth-based version. We employ a straightforward DASH test case for our investigation. The experimental setup comprises three Linux hosts: a DASH server, a video client, and a load generator. These hosts are interconnected via a Tofino 2 switch offering a throughput capacity of 25 Gbps, as depicted in Figure \ref{fig:setup}.

\begin{figure}[ht]
    \centering
    \includegraphics[width=.8\columnwidth]{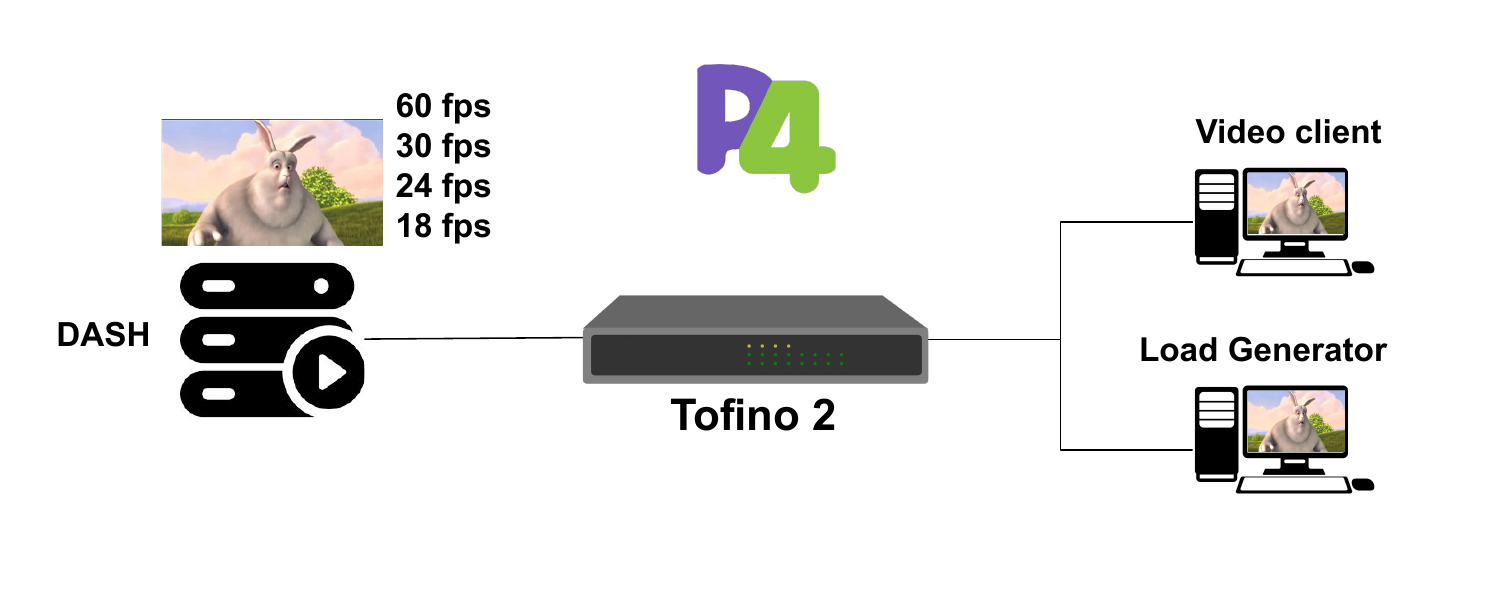}
    \caption{DASH experiment.}
    \label{fig:dash}
\end{figure}

The DASH server houses the Big Buck Bunny video, available in four different quality levels: 60, 30, 24, and 18 frames per second (FPS). The video client possesses the capability to dynamically select from these quality levels based on the prevailing traffic conditions within the network. In instances of elevated network congestion, the client opts for lower-resolution video playback. Conversely, during periods of reduced network load, the client selects the highest available video resolution. This dynamic traffic behavior is influenced by the sinusoidal load applied to the testbed, wherein the number of video clients concurrently consuming the video varies cyclically between 100 and 150 instances.

The load generator generates requests according to a Poisson process, and the arrival rate is modulated by a sinusoidal function as defined in Equation \ref{eq:sinusoid}. In this equation, $ A $ denotes the amplitude, $ F $ represents the frequency, and $ \lambda $ signifies the phase, measured in radians.

\begin{equation}
\textit{$f(y) = A\sin(F + \lambda)$}
\label{eq:sinusoid}
\end{equation}

The video client and load generator share the same output queue in the switch. We set the bandwidth (using port shaping) to 100 Mbps as it is the global average broadband speed \cite{Cisco:2021}.

We conducted an evaluation of various AQM strategies, including iRED, P4-CoDel, and PI2, encompassing measurements at application levels. We examined the FPS rendered by the video client and the size of the local buffer employed for storing and playing forthcoming video frames.

\begin{figure}[ht]
    \centering
    \subfigure[FPS]{
        \includegraphics[width=.45\columnwidth]{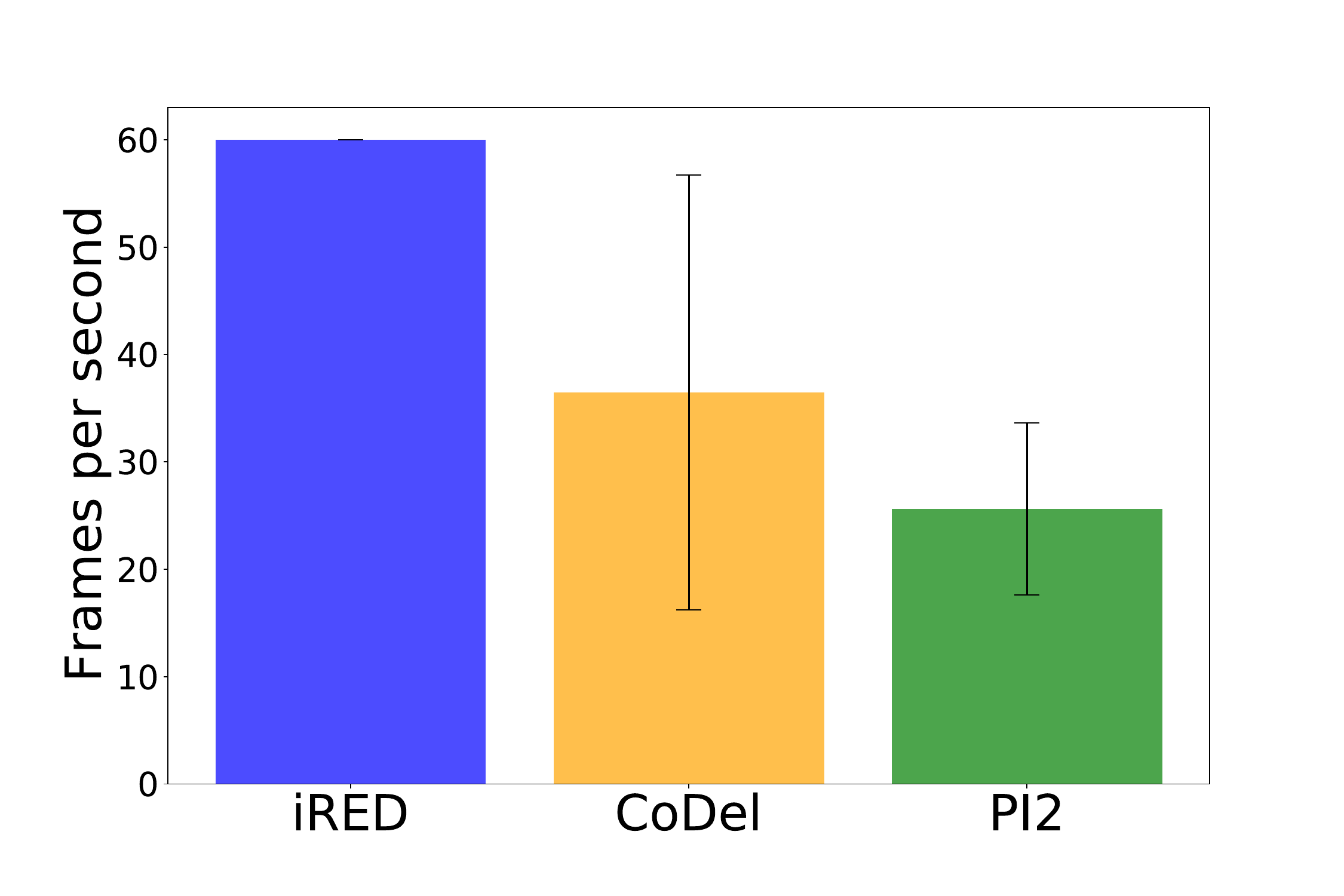}
        \label{fig:FPS}
    }
    \subfigure[Buffer]{
        \includegraphics[width=.45\columnwidth]{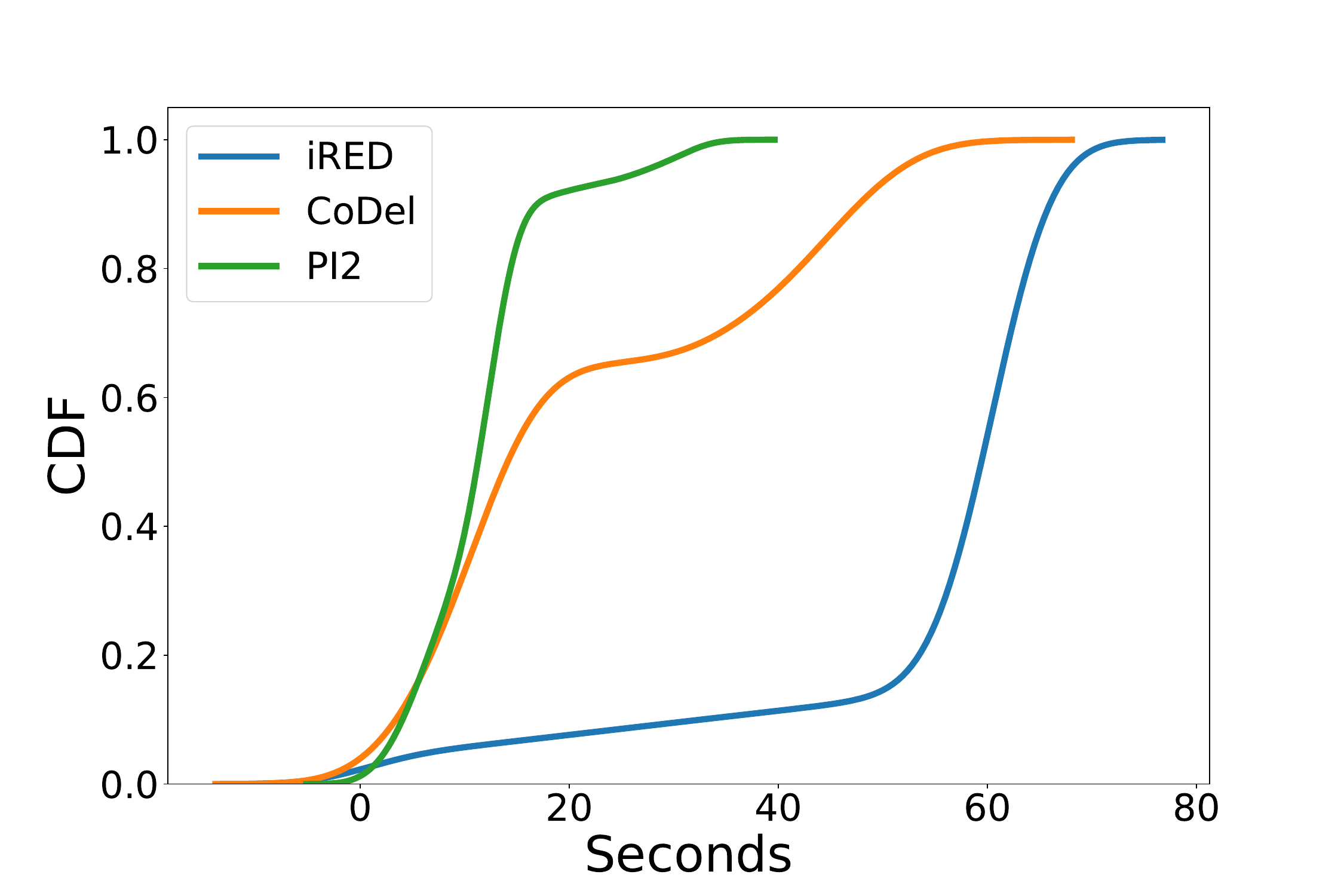}
        \label{fig:buffer}
    }
    \caption{DASH Results. iRED improves the DASH Quality of Service.}
    \label{fig:dash-results}
\end{figure}

Figure \ref{fig:FPS} displays the FPS average achieved by the video client for each AQM approach, while Figure \ref{fig:buffer} presents the cumulative distribution function (CDF) of the remaining buffer duration (in seconds) within the video player. It is evident from the results that iRED optimizes both FPS and the available time in the local buffer for video playback. In light of these findings, it is noteworthy that iRED outperforms P4-CoDel by a factor of 1.64x and PI2 by a factor of 2.34x in terms of maximizing FPS. Regarding the video player buffer, our evaluation shows that iRED allows a filling up to 2.57x compared to P4-CoDel and 4.77x compared to PI2.

Our understanding is that iRED has an advantage for latency-sensitive applications due to packet drops in the Ingress block, which minimizes the waste of switch resources. This is in contrast to the TNA implementations of PI2 and P4-CoDel, where packet drops occur on the egress, potentially resulting in less efficient resource utilization. Additionally, the mechanism for discarding packets in the future has a dual impact. First, it ensures that packets experiencing delay are not immediately discarded but are forwarded to their final destination (the video client). Second, this introduces a subtle delay in signaling congestion at the sender. This delay helps to further smooth out TCP's bursty traffic patterns, making iRED particularly effective in maintaining network stability and reducing congestion-induced fluctuations.

%its utilization of the EWMA of the queue depth. This calculation effectively mitigates the impact of burst traffic patterns, making iRED better suited to maintaining low latency in such scenarios. Additionally, it's worth noting that iRED employs 

\section{Conclusions} \label{sec:conclusions}

Traditionally, AQMs are countermeasures to alleviate transient congestion, aiming to maintain high throughput and low delay in queues. In essence, they detect incipient network congestion, e.g., based on the queue length, and provide congestion notification to end-hosts by dropping/marking packets, allowing them to back off before queue overflow and sustained packet loss occurs.

In this work, we presented iRED, a disaggregated P4-AQM fully implemented and tested in programmable data plane hardware. iRED is a deployment of the RED algorithm in a Tofino2 P4-switch that supports the L4S framework, capable of dropping (Classic) or marking (Scalable) packets using the coupling mechanism. Moreover, we created a modified version of iRED using the very new feature of Tofino2 (Ghost thread), which allows us to consult queue depth information at the Ingress block. In addition, we clarify the AQM operations (decision and action) and the Egress drop problem for the state-of-the-art AQMs implemented in the PDP hardware, showcasing the primary wasted resources associated with this approach.

Based on our results, we confirm that the decision to drop or not a given packet should be kept at the Egress block, and then, when needed, send very small notification packets (minimum overhead) to the Ingress block. Using this design, the device can significantly save resources in terms of memory usage, queue delay, clock cycles, and power consumption.

%We believe the decision should be kept from the Egress block, as only notification packets (minimum overhead) need to be forwarded to the Ingress block. Using this design, the device can save resources in terms of memory usage, queue delay, clock cycles, and power consumption. \textcolor{red}{In complex networks, maybe packets are discarded by routers very close to the destination after wasting resources along the path. A next step could be, for example, to recirculate the 48 bytes packet upstream the path, to the first router on the path, saving the entire path resources.}

%\bmhead{Acknowledgments}

%%===========================================================================================%%
%% If you are submitting to one of the Nature Portfolio journals, using the eJP submission   %%
%% system, please include the references within the manuscript file itself. You may do this  %%
%% by copying the reference list from your .bbl file, paste it into the main manuscript .tex %%
%% file, and delete the associated \verb+\bibliography+ commands.                            %%
%%===========================================================================================%%
\bibliography{sn-article}% common bib file
%% if required, the content of .bbl file can be included here once bbl is generated
%%\input sn-article.bbl
\end{document}